\documentclass[12pt,hidelinks]{article}
\pdfoutput=1
\usepackage[body={17.5cm, 22cm},right=2cm]{geometry}
\usepackage[table]{xcolor}
\usepackage{color}
\usepackage{amssymb,graphicx}
\usepackage{tikz-feynman} 
\usepackage{amsmath}
\usepackage{epsfig}
\usepackage{lmodern}
\usepackage[normalem]{ulem}
\usepackage{bm}
\usepackage[bookmarks=true,pdfborder={0 0 0}]{hyperref} 
\usepackage{feynmp-auto}
\usepackage{tabularx}
\usepackage{slashed}

\usepackage{authblk}

\usepackage{caption}
\usepackage{subcaption}

\usepackage{cite} 
\usepackage{hyperref}

\newcommand{\sun}{\odot}

\newcommand{\km}{{\, {\rm km}}}

\newcommand{\MeV}{{\, {\rm MeV}}}

\newcommand{\TeV}{{\, {\rm TeV}}}

\newcommand{\gram}{{\, {\rm g}}}
\newcommand{\erg}{{\, {\rm erg}}}

\def\beq{\begin{equation}}
\def\eeq{\end{equation}}
\def\bea{\begin{eqnarray}}
\def\eea{\end{eqnarray}}
\def\bitem{\begin{itemize}}
	\def\eitem{\end{itemize}}
\newcommand{\bec}{\begin{center}}
	\newcommand{\eec}{\end{center}}
\newcommand{\ba}{\begin{array}}
	\newcommand{\ea}{\end{array}}

\DeclareMathOperator{\Rea}{Re}
\DeclareMathOperator{\Ima}{Im}

\usepackage{float}
\allowdisplaybreaks

\begin{document}

\setcounter{page}{0}
\thispagestyle{empty}

\parskip 3pt

\font\mini=cmr10 at 2pt

\begingroup
\renewcommand{\thefootnote}{\alph{footnote}} 
\setcounter{footnote}{0} 

\begin{titlepage}
\noindent \makebox[15cm][l]{\footnotesize \hspace*{-.2cm} }  \\  [-1mm]
~\vspace{2cm}

\begin{center}

{\LARGE \bf Stellar cooling limits on KK gravitons  \\ \vspace{0.4cm} and dark dimensions}

\vspace{0.9cm}

{\large
Edward Hardy, Anton Sokolov, Henry Stubbs\,\footnote{\url{henry.stubbs@physics.ox.ac.uk} (corresponding author)}}
\\
\vspace{.6cm}
{\normalsize { \sl Rudolf Peierls Centre for Theoretical Physics, University of Oxford, \\ Parks Road, Oxford OX1 3PU, UK}}

\end{center}

\vspace{.1cm}
\begin{abstract}
We revisit cooling bounds on light Kaluza-Klein (KK) gravitons, as arise in the dark dimension scenario, considering red giants, neutron stars, and supernovae. In addition to bremsstrahlung, we account for two novel production channels: resonant mixing with the in-medium photon and a pion-induced process in supernovae. The strongest limits arise from SN~1987A, with the emissivity from the pion process exceeding that from bremsstrahlung by a factor of a few. Given present uncertainties, we obtain a bound on the KK mass scale of $m_{\rm KK}\gtrsim 0.6\,{\rm eV}$ ($\gtrsim 500\,{\rm eV}$) for $2$ ($3$) extra dimensions. Improved understanding of the properties of pions in supernovae could strengthen these limits to roughly ${\rm eV}$ (${\rm keV}$). For $1$ extra dimension, the bounds are weaker than those from laboratory searches. We also show that constraints from KK graviton decays to Standard Model particles are less stringent than the cooling bounds if there is  KK number violation at the level typically assumed in the dark dimension scenario, although these bounds could be strengthened by future observations.

\end{abstract}

\end{titlepage}

\endgroup 

\setcounter{footnote}{0}
\renewcommand{\thefootnote}{\arabic{footnote}}

{\fontsize{11}{10.8}
\tableofcontents
}

\section{Introduction}

The idea that the Universe might contain relatively large extra dimensions with the Standard Model (SM) fields confined to a brane was originally motivated by the Electroweak hierarchy problem \cite{Arkani-Hamed:1998jmv,Antoniadis:1998ig,Arkani-Hamed:1998sfv}. In this context, the fundamental higher-dimensional Planck scale $M_D$, which is the UV cutoff of the theory, is assumed to be close to the $\TeV$ scale. The radius $R$ of the extra dimensions must be such that the 4-dimensional Planck scale $M_{\rm Pl}$ takes its measured value, with
\beq \label{eq:1}
M_{\rm Pl}^2 \simeq \left(2\pi R\right)^{n} M_D^{n+2} ~, 
\eeq
where $n$ is the number of large extra dimensions (throughout, we assume that these are approximately toroidally compactified with a common radius $R$, but our results can easily be adapted to other setups, e.g. \cite{Lykken:1999ms}). Although theoretically appealing, strong lower bounds on $M_D$ from collider searches, e.g. for signals of KK gravitons, have tempered the viability of this as a solution to the hierarchy problem.

The possibility of large extra dimensions has recently been revived in Ref.~\cite{Montero:2022prj} motivated by the value of the cosmological constant and string theory ``swampland'' conjectures \cite{Vafa:2005ui}. In this proposal, known as the  \emph{dark dimension}, $R$ is assumed to be of order ${\rm eV}^{-1}$. For $n=1$ and $2$, the corresponding $M_D$ are around $10^{10}\,{\rm GeV}$ and $10^5\,{\rm GeV}$ respectively, which are allowed by high energy particle collider constraints, while larger $n$ are already excluded \cite{ParticleDataGroup:2024cfk}. 
Crucial open questions remain concerning whether a full string theory construction of a  dark dimension can be consistent with existing observational and cosmological constraints, for example on light scalar moduli fields that might typically accompany it. Moreover, further theoretical study of the swampland arguments that led to the proposal is essential \cite{Branchina:2023ogv}. 
Nevertheless, given the scarcity of ideas to understand the value of the cosmological constant, it is worthwhile to analyse current constraints on dark dimensions and the prospects for future improvements or discovery.

Among the possible signals of large extra dimension scenarios, the experimental and observational implications of KK gravitons are especially interesting because these particles are unavoidable. 
For $n=1,2$, with $M_D$ well above collider scales, the laboratory bounds come from searches for deviations from Newtonian gravity at short distances  \cite{Lee:2020zjt} due to the lightest KK gravitons in the tower. These constrain  $m_{\rm KK}\gtrsim 5\,{\rm meV}$, with a mild dependence on $n$ \cite{Adelberger:2003zx}. Meanwhile, astrophysical constraints on KK gravitons have been studied extensively since the original proposal of large extra dimensions \cite{Arkani-Hamed:1998sfv,Barger:1999jf,Cullen:1999hc,Hanhart:2000er,Hanhart:2001fx,Hannestad:2001jv,Hannestad:2001xi,Hannestad:2003yd} and give the strongest bounds for $n=2$ and $3$.

The most model-independent astrophysical bounds  
are due to ``cooling'', in which KK gravitons 
produced in a hot stellar core escape, carrying away energy and thereby altering the evolution of a star (see e.g. \cite{Raffelt:1996wa,Caputo:2024oqc} for comprehensive reviews). The strongest known cooling limits on large extra dimensions arise from the neutrino burst observed from the (presumed) protoneutron star in SN~1987A. 
These are weaker than the laboratory bounds for $n=1$, while for $n\geq 2$ they are stronger due to the larger number of energetically accessible KK states. Existing SN~1987A cooling constraints allow either one or two extra dimensions of the size suggested in the dark dimension scenario \cite{Hannestad:2003yd}.

Even stronger constraints arise in specific theories; for example, if the KK gravitons decay only to SM particles then  observations of the heating of neutron stars  lead to $m_{\rm KK}\gtrsim 3\,{\rm eV}$ for $n=2$ (the bounds for $n=1$ are again weaker than laboratory constraints) \cite{Hannestad:2001jv,Hannestad:2001xi,Hannestad:2003yd,Casse:2003pj}. However, as noted in the original work \cite{Hannestad:2001jv}, and further analysed in \cite{Mohapatra:2003ah,Anchordoqui:2025nmb}, these limits do not apply if the KK gravitons dominantly decay to lighter KK gravitons due to a violation of KK number.

In this paper, we revisit the cooling constraints on KK gravitons from a wide range of astrophysical objects. In addition to the well-studied bremsstrahlung emission, we consider two new production processes. The first is  resonant mixing of the in-medium photon (plasmon) with KK gravitons, which occurs when the dispersion relations of the two coincide. Similar production has previously led to strong limits on new light scalars from red giants and horizontal branch stars, and it is also relevant in supernovae  \cite{Hardy:2016kme,Hardy:2024gwy} (it may also play a role for white dwarfs, although we leave these to future work). The second is the production of KK gravitons in supernova from pions, specifically $\pi^- p\rightarrow n h $, where $h$ is the KK graviton. For axions, a similar process has been found to be slightly more important than bremsstrahlung in supernova \cite{Carenza:2020cis}, although there are uncertainties on the abundance and properties of pions in protoneutron stars.

Finally, we quantify the impact of the KK graviton decays to other KK gravitons on the model-dependent limits from decays to SM particles. This sets an upper bound on the KK number violation if cooling limits are to be dominant. 
Our work is complementary to the recent detailed study of Ref.~\cite{Anchordoqui:2025nmb}, which analysed collider and cosmological constraints on dark dimension scenarios (we note that this reference also considered the impact of intra-tower KK graviton decays; our result for the decay rate differs from theirs, although we reach similar conclusions) and Ref.~\cite{Law-Smith:2023czn}, which considers constraints from the decays of dark matter KK gravitons to the SM (focusing on the $n=1$ case). Other recent works studying the phenomenology of the dark dimension scenario include Refs.~\cite{Anchordoqui:2022txe,Anchordoqui:2022svl,Anchordoqui:2022tgp,Blumenhagen:2022zzw,Anchordoqui:2022ejw,Cui:2023wzo,Anchordoqui:2023oqm,Anchordoqui:2023tln,Anchordoqui:2023wkm,Law-Smith:2023czn,Anchordoqui:2023laz,Burgess:2023pnk,Dvali:2023zww,Anchordoqui:2024akj,Anchordoqui:2024dxu,Gendler:2024gdo,Basile:2024lcz,Casas:2024oak,Heckman:2024trz,Gross:2024zdh,Li:2024jko,Anchordoqui:2024xvl,Bedroya:2025fwh,Antoniadis:2025rck}.

{\it Conventions:} 
we define $\kappa=\sqrt{32\pi G_N}$, where $G_N$ is Newton's constant, $R$ is the radius of the extra dimension such that the mass of the lightest KK mode $m_{\rm KK}=1/R$, and our spin-2 polarization vectors are normalised such that $\epsilon^{s,\mu\nu} \epsilon^{s'*}_{\mu \nu}= \delta^{ss'}$ (which matches Ref.~\cite{Gleisberg:2003ue} and differs from version 4 of Ref.~\cite{Han:1998sg} by some factors of $2$ at intermediate steps).

The structure of this paper is as follows. In Section~\ref{sec:prod} we calculate the rate of resonant KK graviton production, which is relevant to red giants and supernovae. In Section~\ref{sec:RG}, we analyse the limits from the tip of the red giant branch,  
and in Section~\ref{sec:NS} we obtain the bounds from old neutron stars. In Section~\ref{sec: supernova} we study the bounds from SN~1987A, and in Section~\ref{sec:decay} we analyse the impact of decays within the KK graviton tower. We conclude and discuss possible future improvements in Section~\ref{sec:conc}. Additional technical details are given in Appendices.

\section{Thermal production and KK graviton mixing \label{sec:prod}}

The production rate of a weakly coupled boson $x$ in a thermal medium can be shown to be~\cite{Laine:2016hma,Hardy:2024gwy}
\begin{equation}\label{eqn:full production rate}
    \frac{dN}{dV dt} = -g\int\frac{d^3 k}{(2\pi)^3}\frac{f_B(\omega)}{\omega}\Ima\left[\Pi_{xx}-\frac{(\Pi_{\gamma x})^2}{\Pi_{\gamma\gamma}-m_x^2}\right] +\mathcal{O}(\lambda^4),
\end{equation}
where $\Pi_{\gamma \gamma}$ ($\Pi_{xx}$) denotes the photon's (weakly coupled boson's) thermal self-energy~\cite{Bellac:2011kqa,Laine:2016hma}, and $\Pi_{\gamma x}$ is the 1PI mixing between them. All Lorentz indices have been contracted with appropriate polarization vectors and $g$ is the number of states for each polarization.\footnote{Later, a superscript $L/T$ will be used to indicate what polarization vectors the self-energies have been contracted with. $g=1$ for longitudinal mixing and $g=2$ for transverse mixing.} The (assumed small) coupling between the boson and the SM is denoted $\lambda$, and $\omega = \sqrt{k^2+m_x^2}$.

If the mass of the new boson, $m_x$, exceeds the plasma frequency, $\omega_p$, which is the typical scale of $\Pi_{\gamma\gamma}$, the second term in the square bracket in Eq.~\eqref{eqn:full production rate} can be neglected. 
The first term contains all of the typical contributions; depending on the interactions of $x$, these include electron annihilation, Compton-like emission, and bremsstrahlung. This can be evaluated in various ways, for example a full finite-temperature field theory calculation using cutting rules~\cite{Kobes:1985kc,Bedaque:1996af,Gelis:1997zv}, or by other systematic methods~\cite{Brahma:2025wos}. However, we will employ the simple and commonly used approach of thermally averaging the vacuum interaction rates for these processes  (although this may miss interesting thermal effects such as the effective nucleon mass in supernovae).

For light beyond Standard Model (BSM) particles, the second term in Eq.~\eqref{eqn:full production rate} may be significant if the BSM particle has quantum numbers such that $\Pi_{\gamma x}$ can be non-zero. For new particles with interactions such that $\Pi_{xx}$ is related to $\Pi_{\gamma x}$ (e.g. dark photons for which $\Pi_{xx}=\lambda \Pi_{x\gamma}=\lambda^2 \Pi_{\gamma\gamma}$) cancellations between the two terms can be important. However, this does not typically occur for more general new particle interactions (including for KK gravitons) and each term can be considered separately. Then, expanding the self-energies into real and imaginary parts, and assuming that $\Rea[\Pi_{\gamma x}]\gg\Ima[\Pi_{\gamma x}]$, the second term becomes
\begin{equation}
    \frac{dN}{dV dt} = g\int\frac{d^3k}{(2\pi)^3}\frac{f_B(\omega)}{\omega}\Rea[\Pi_{\gamma x}]^2\frac{-\Ima[\Pi_{\gamma\gamma}]}{(\Rea[\Pi_{\gamma \gamma}]-m_x^2)^2-\Ima[\Pi_{\gamma\gamma}]^2}.
\end{equation} 
 Provided that the lifetime of the photon in the medium is long, such that $\Ima[\Pi_{\gamma\gamma}]$ is small, this contribution has a sharp resonance when $m_x^2 = \Rea[\Pi_{\gamma\gamma}]$,  which might dominate the total production rate. 
 
 In the narrow resonance regime, we may make a delta function approximation to give
 \begin{equation}
     \frac{dN}{dV dt} = \frac{g}{2\pi}\int d\omega \frac{k}{e^{\omega/T}-1} \Rea[\Pi_{\gamma x}]^2 \delta(\Rea[\Pi_{\gamma\gamma}]-m_x^2),
 \end{equation}
 which can be evaluated to give the energy production rate
 \begin{equation} \label{eq:epro}
     \frac{dQ}{dV dt} = \frac{g}{2\pi}\frac{k_*\omega_*}{e^{\omega_*/T}-1}\Rea[\Pi_{\gamma x}(\omega_*)]^2\left|\frac{d\Rea[\Pi_{\gamma\gamma}(\omega)]}{d\omega}\right|^{-1}_{\omega = \omega_*},
 \end{equation}
where $\omega_*$ and $k_*$ are the particle's energy and momentum evaluated on-resonance.\footnote{The self-energies have been written as a function of $\omega$ only, since the dependence on $k$ is fixed from the beginning by the boson's on-shell condition $\omega^2 = k^2+m_x^2$.} 
Using the standard expressions for $\Rea[\Pi_{\gamma\gamma}]$ \cite{Braaten:1993jw}, we have
\begin{align}
    \left.\frac{d\Rea[\Pi_{\gamma\gamma}^L(\omega)]}{d\omega}\right\vert_{\omega = \omega_*} &= -\omega_*\frac{m_x^2}{k_*^2}
    \left(2+\frac{m_x^2}{\omega_*^2}\frac{(\omega_*^2-k_*^2v_0^2)-3\omega_p^2}{\omega_*^2-k_*^2v_0^2}\right), 
    \\ \label{Eqn. Transverse derivative factor}
    \left.\frac{d\Rea[\Pi_{\gamma\gamma}^T(\omega)]}{d\omega}\right \vert_{\omega = \omega_*} &= -\omega_*\frac{m_x^2}{k_*^2}\left(\frac{m_x^2}{\omega_*^2}-\frac{3\omega_p^2-2m_x^2}{\omega_*^2-k_*^2 v_0^2}\right), 
\end{align}
where $v_0$ is the typical velocity of electrons that contribute to the photon dispersion, equal to the Fermi velocity for degenerate plasmas.\footnote{For $v_0=1$, this reproduces the result shown in Ref.~\cite{Hardy:2024gwy}. For $v_0\rightarrow0$ the transverse case has $m_x\rightarrow\omega_p$ on resonance, and so Eq.~\eqref{Eqn. Transverse derivative factor} goes to zero, which causes the production rate per unit volume to diverge. Physically, this is because in the classical limit the transverse resonance occurs in an infinitesimally thin spatial region.}

\begin{figure}[htbp]
  \begin{minipage}{0.98\textwidth}
    \centering
    \includegraphics[width=0.65\textwidth]{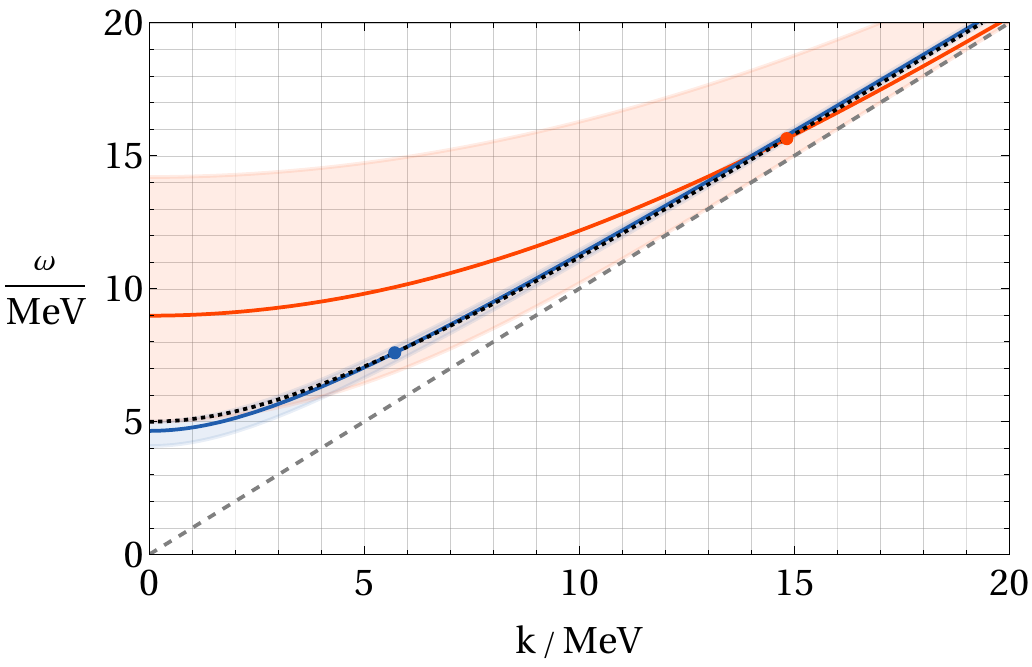}
    \caption{
    The energy-momentum relation of a KK graviton of mass $m_h=5\MeV$  (dashed black). $m_h$ determines the plasma frequencies $\omega_p$, and hence the locations in a star, where resonant production occurs. For the longitudinal mode, there is a resonance  whenever $\omega_p \geq m_h$. 
 For transverse modes, there is a resonance when $\omega_p\in \left[\sqrt{2/3}m_h, m_h\right]$. The regions spanned by all such longitudinal (transverse) photon dispersions in a typical protoneutron star are shown in shaded red (blue) (specifically the $20M_\sun$ progenitor model discussed in Section~\ref{sec: supernova}, which has  maximum $\omega_p\simeq 14\,{\rm MeV}$). Representative dispersion curves are plotted in bold red (blue), and the resonance points where they intersect the KK graviton curve are marked with solid dots. }
    \label{fig:dispersions1}
  \end{minipage}\hfill
  \vspace{0.5cm}
  \begin{minipage}{0.98\textwidth}
    \centering
    \includegraphics[width=0.65\textwidth]{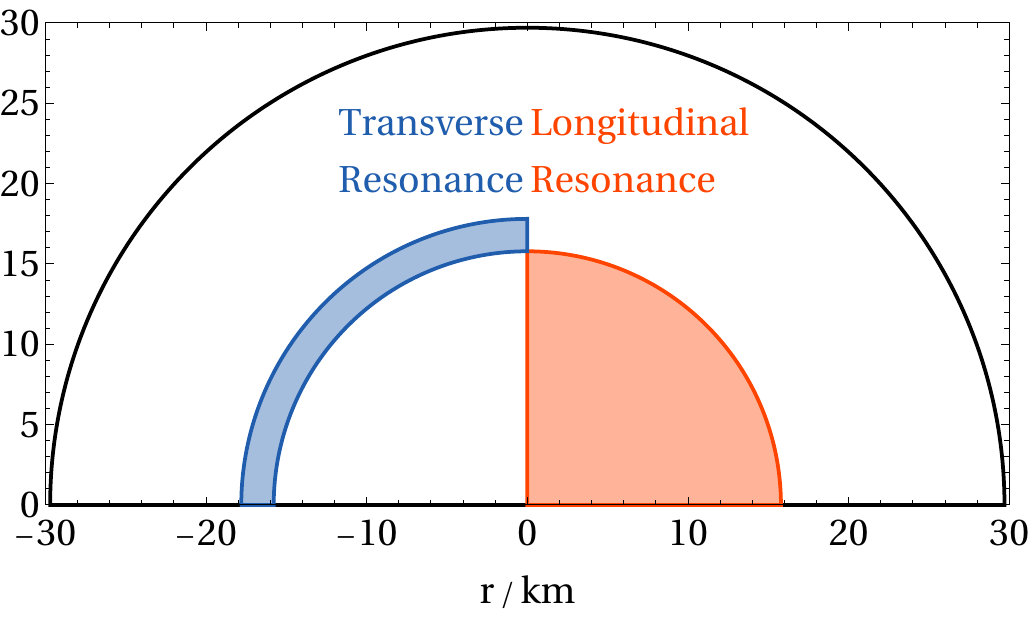}
    \caption{The regions within a protoneutron star core where the different resonances can occur for a KK graviton of mass $5\MeV$, using the same model of the star as in Figure~\ref{fig:dispersions1}. A longitudinal resonance occurs whenever $m_x\leq\omega_p(r)$, and therefore throughout the region $0<r<r_1$ where $\omega_p(r_1)=m_x$ (shaded red). In the limit that the electrons are relativistic, a transverse resonance only occurs if $\omega_p(r)\leq m_x \leq \sqrt{3/2} \omega_p(r)$. This therefore takes place in some shell $r_1<r<r_2$ where $\omega_p(r_2)=\sqrt{2/3}m_x$ (shaded blue). Here $r_1\approx16\km$ and $r_2\approx18\km$.}
    \label{fig:dispersions2}
  \end{minipage}
\end{figure}

Inside a star, $\omega_p$ typically decreases with increasing radial distance $r$.  For longitudinal photons, a resonance always occurs if $m_x<\omega_p$, corresponding to all $r$ up to some maximum radius where $\omega_p(r)=m_x$. For transverse photons there is only a resonance if $m_x$ is in some narrow window above $\omega_p$ meaning that, if $m_x$ lies roughly between the smallest and largest values of $\omega_p$ in the star, there is a spatially thin shell where the resonance can occur. The width of this shell depends on the typical velocity of the electrons in the medium, and in the non-relativistic limit it can be treated as a delta function. This is illustrated in Figures~\ref{fig:dispersions1} and \ref{fig:dispersions2}, which show the  dispersion relations crossing and the relevant spatial regions. The situation is more complicated in the case of KK gravitons, because there is an almost continuous spectrum of massive spin-2 bosons with both longitudinal and transverse polarization states. Both can be produced resonantly, and the range of masses means that resonant production can occur anywhere within the star.

\begin{fmffile}{self_energies}
\begin{table}
    \centering    
\begin{tabularx}{0.9\textwidth} { 
  | >{\centering\arraybackslash}X 
  | >{\centering\arraybackslash}X 
  | >{\centering\arraybackslash}X | }
    \hline & &  \\
    %
    %
     \begin{fmfgraph*}(110,100)
    \fmfleft{i}
    \fmfright{o}
    \fmf{photon}{i,v1}
    \fmf{dbl_wiggly}{v2,o}
    \fmf{fermion,right,tension=0.5}{v1,v2}
    \fmf{fermion,right,tension=0.5}{v2,v1}
\end{fmfgraph*}   
     &
     \begin{fmfgraph*}(110,100)
    \fmfleft{i}
    \fmfright{o}
    \fmftop{t}
    \fmf{photon,tension=100}{i,v1}
    \fmf{dbl_wiggly,tension=100}{v1,o}
    \fmf{photon}{v1,v2}
    \fmf{fermion,right,tension=0.2}{v2,t}
    \fmf{fermion,right,tension=0.2}{t,v2}    
\end{fmfgraph*}
     &
    \begin{fmfgraph*}(110,100)
    \fmfleft{i}
    \fmfright{o}
    \fmftop{t1,t2,A,t3,t4}
    \fmf{photon,tension=100}{i,v1}
    \fmf{dbl_wiggly,tension=100}{v1,o}
    \fmf{fermion}{v1,t3}
    \fmf{fermion}{t3,t2}
    \fmf{fermion}{t2,v1}
\end{fmfgraph*}
    \\[-20pt]
     \hline & &  \\
    %
    %
    \begin{fmfgraph*}(110,60)
    \fmfleft{i2,i1}
    \fmfright{o2,o1}
    \fmf{fermion}{i1,v1}
    \fmf{fermion}{v1,v2}
    \fmf{fermion}{v2,o1}
    \fmf{photon}{i2,v1}
    \fmf{dbl_wiggly}{v2,o2}
    \end{fmfgraph*}   
    &
    \begin{fmfgraph*}(110,60)
    \fmfleft{i1,i2}
    \fmfright{o1,o2}
    \fmf{photon}{i1,v1}
    \fmf{dbl_wiggly}{v1,o1}
    \fmf{photon}{v1,v2}
    \fmf{fermion}{i2,v2}
    \fmf{fermion}{v2,o2}
\end{fmfgraph*}   &
     \begin{fmfgraph*}(110,60)
    \fmfleft{i1,i2}
    \fmfright{o1,o2}
    \fmf{photon}{i1,v1}
    \fmf{dbl_wiggly}{v1,o1}
    \fmf{fermion}{i2,v1}
    \fmf{fermion}{v1,o2}
\end{fmfgraph*}   
    \\[10px]
     \hline & &
     \\[-10px]
     s and u channel scattering (only the s channel is shown)&
     t channel scattering&
     Four-point contact diagram
     \\[15px]
    \hline
\end{tabularx}

\caption{The Feynman diagrams contributing to the photon--KK graviton mixing $\Pi_{\gamma h}$, and the cut diagrams showing the correspondence with various scattering channels. Single wavy lines are photons, and double wavy lines are gravitons. The fermion loops contain contributions from all species present in the plasma.}
    \label{tab:SelfEnergyDiagramTable}
\end{table}
\end{fmffile}

To determine the resonant production rate we need to evaluate $\Rea[\Pi_{\gamma x}(\omega_*)]$. In the case of KK gravitons, $h$, the diagrams contributing to this are shown in Table~\ref{tab:SelfEnergyDiagramTable}, together with the corresponding forward scattering processes (for the connection see, e.g., Ref.~\cite{Braaten:1993jw}).

We calculate the photon--KK graviton plasma mixing in various regimes in Appendix~\ref{Appendix: mixing calculation}. Due to charge-conjugation (C) symmetry violation being required for the mixing, this vanishes if the plasma has equal numbers of particles and antiparticles, as in the case of scalar-photon mixing~\cite{Hardy:2016kme}. However, unlike in previously studied cases, the non-relativistic mixing is suppressed by a factor of $T/m_f$, where $m_f$ is the mass of the fermion in the medium, so it vanishes in the $T\rightarrow0$ limit.
The physical justification for this can be seen from the classical equations of motion, in the limit where both the electrons and ions of the medium begin at rest. An incoming electromagnetic wave displaces the electrons and ions in opposite directions. Usually, the motion of the ions is neglected~\cite{Hardy:2016kme,Dubovsky:2015cca} because their displacement is suppressed by a factor of $m_e/m_i$. However in our case, the coupling to gravity is larger for the ions by a factor $m_i/m_e$. The two contributions to the source term for the gravitons then cancel out, leading to no mixing. 
The mixing is also suppressed in the limit where all particles are ultra-relativistic; this time by a factor $m_f^2/T^2$, and vanishes in the limit of massless fermions.\footnote{This cannot be seen directly in the final results of Appendix~\ref{Appendix: mixing calculation}, as the function $\Bar{H}$ has to be expanded to a higher order to get a non-zero result.}

Resonant production may therefore play a role in astrophysical plasmas with a mixture of non-relativistic and relativistic particle species, provided $\omega_p$ is not much larger than the temperature, otherwise the Boltzmann suppression in Eq.~\eqref{eq:epro} is substantial. Such systems include red giants, white dwarfs, and the cores of supernovae. For older neutron stars $\omega_p\gg T$, while in horizontal branch stars and in main sequence stars such as the Sun the medium is close to non-relativistic, so resonant production is negligible in these environments. 
For dark photons, resonant conversion is also active in the early-Universe plasma, which has been thought to play a role in cosmological constraints, although this has recently been reconsidered \cite{Hook:2025pbn}. For KK gravitons, however, the near $C$-symmetry of this environment suppresses such processes, rendering them irrelevant.

In addition to resonant production, various processes including bremsstrahlung radiation and Compton-like emission can produce KK gravitons through the first term in Eq.~\eqref{eqn:full production rate}. Their relative importance depends on the particular astrophysical system, so we defer an analysis to later sections.

\section{Tip of the red giant branch \label{sec:RG}}

As a red giant (RG) uses up its hydrogen fuel, the mass of its degenerate helium core and hence its luminosity increases until the core becomes hot and dense enough to begin to fuse helium into carbon. Due to the degeneracy of the core, the associated increase in temperature leads to a runaway helium fusion via the triple-alpha process  (the so-called helium flash), which liberates a vast amount of energy in a short period of time and ultimately causes the core to expand. The luminosity of the star then decreases, meaning that the moment of helium ignition sets the maximum luminosity of a red giant star, which is known as the tip of the red giant branch (TRGB). In the presence of any additional cooling the core would grow heavier and more luminous before the helium flash can occur, making the resulting TRGB brighter. Since the triple-alpha process has a strong dependence on the properties of the core, the TRGB brightness is a sensitive probe of new hypothetical particles that could be produced in the core and carry energy away~\cite{Raffelt:1996wa}. The main SM cooling mechanism for the core is plasmon decay to neutrinos, with an average energy loss rate just before the helium ignition of $\epsilon_\nu \approx 4\erg\gram^{-1}\sec^{-1}$ \cite{Straniero:2020iyi}. Recent investigations of the TRGB in Globular Clusters (GCs) indicate that the average novel energy loss can be constrained by data at the level $\epsilon_x\lesssim (10^{-1} \div 1) \, \epsilon_{\nu}$~\cite{Viaux:2013lha,Capozzi:2020cbu,straniero2018axionelectroncouplingrgbtip,Straniero:2020iyi,Troitsky:2024keu,Caputo:2024oqc}. Here, we assume a conservative constraint $\epsilon_x < \epsilon_{\nu}$.

To estimate the rate of emission of KK gravitons from the RG core at helium ignition, we generate profiles of thermodynamic characteristics inside the core for different possible values of initial star metallicities $Z_{\text{ini}} = \left\lbrace 10^{-4},\, 10^{-3},\, 6\times 10^{-3} \right\rbrace$, with the initial helium abundance $Y_{\text{ini}}=0.25$ and the age at the TRGB $\simeq \!13$~Gyr. These particular choices of parameters correspond to a recent detailed analysis of the relevant GCs in Ref.~\cite{Straniero:2020iyi}. Specifically, we evolve the one-dimensional stellar models with the MESA code \cite{Paxton:2010ji} starting from the zero-age main sequence model with the chemical composition prescribed above and the initial mass corresponding to the above TRGB age, until the moment when helium-burning luminosity reaches $L_{\text{He}} > 10^2 L_{\odot}$. The resulting  temperature, density and plasma frequency profiles are shown in Fig.~\ref{Fig: RG profiles} in Appendix~\ref{app:RG}.

In the past, continuum production channels of the KK gravitons inside a RG core have been considered. The most important of these are plasmon annihilation, gravi-Compton-Primakoff scattering and gravi-bremsstrahlung. In Ref.~\cite{Barger:1999jf} the gravi-bremsstrahlung process was found to dominate over the other two. However, in this calculation the plasma was assumed to be non-degenerate, which is not a good approximation inside the highly degenerate RG core at the time of helium ignition, with such an assumption leading to an overestimation of the gravi-bremsstrahlung rate by several orders of magnitude~\cite{Hansen:2015lqa}. 

We find that resonant production results in an energy loss rate that is similar to the gravi-bremsstrahlung \textit{before} the degeneracy correction for the latter is taken into account, and therefore we conclude that resonant emission dominates over continuum production.

Our calculations show that longitudinal resonant emission surpasses the transverse one only for lower KK masses, and that the full emission from the tower is dominated by the transverse resonance. With the resonant condition applied, the main contribution to the mixing self-energy for transverse polarizations is (see Eq.~\eqref{eqn: T deg mixing simplified} in Appendix~\ref{app:summarymix}):
 \begin{equation}\label{mixpit}
    \Pi_{\gamma h,f}^{T} = \frac{q_f e \kappa}{3 \sqrt{2} \pi^2} p_F^3\frac{m_h}{v_* \omega_*}\left(1-\frac{m_h^2}{\omega_p^2}\right) \, ,
\end{equation}
where $f$ denotes the fermion that leads to the mixing (with $q_f$ its electric charge and $p_F$ its Fermi momentum), $m_h$ is the mass of a particular KK graviton mode, $\omega_*$ is the energy at resonance, and $v_*=k_*/\omega_*$ is the KK graviton velocity where $k_*$ is the KK graviton momentum at resonance. In a RG, the mixing is dominated by degenerate electrons, which have Fermi momentum of about $0.25\,{\rm MeV}$. 
Because the electrons are only mildly relativistic, the resonance occurs only for $m_h \simeq \omega_p$, as follows from the functional form of the self-energy of a transverse plasmon in degenerate plasma~\cite{Braaten:1993jw,Raffelt:1996wa}. Consequently, the factor in the brackets in the right-hand side of Eq.~\eqref{mixpit} is  small $(1-m_h^2/\omega_p^2) \sim 10^{-2}$, which suppresses the rate of KK graviton emission.

The resulting constraints on the size of extra dimensions are given in Table~\ref{tab:red_giant_limits}; evidently the assumed metallicity makes  no appreciable difference. We stress that the bound on the extra energy loss that we impose, $\epsilon_x\lesssim4\erg\gram^{-1}\sec^{-1}$, is quite conservative. For instance, a detailed analysis for the case of axions~\cite{Straniero:2020iyi}, in which axions are included in the stellar evolution, suggests $\epsilon_a\lesssim 0.9\erg\gram^{-1}\sec^{-1}$~\cite{Caputo:2024oqc,Carenza:2024ehj}. Moreover, a recent re-analysis~\cite{Troitsky:2024keu} using data on GCs from the Data Release 3 from the Gaia mission~\cite{Gaia:2016zol,Gaia:2023fqm} yields a further factor of 2 improvement in the constraints on the axion coupling, meaning that the energy loss bound is $\epsilon_a\lesssim 0.2\erg\gram^{-1}\sec^{-1}$. If naively rescaled to the case of extra dimensions, this would imply an improvement to the bounds in Table~\ref{tab:red_giant_limits} of a factor of $20^{1/n}$, where $n$ is the number of extra dimensions. Such a translation may be well-justified, since degenerate electrons in the RG core are good heat conductors, and so one can expect that the constraint is determined primarily by the total energy loss rather than by a detailed emission profile. Furthermore, we expect that even this constraint might be improved in the near future given the abundance of recent observational data on GCs, e.g. from the Gaia mission \cite{Gaia:2016zol,Li_2023}.

\begin{table}[hbt!]
    \centering
    \def\arraystretch{1.3}
        \begin{tabular}{|c|c|c|c|}
    \hline
         metallicity & $n=1$  & $n=2$  & $n=3$  \\
         \hline\hline
        $10^{-4}$ & $1.0\times10^{4}$& $2.1\times10^{-4}$ & $6.5\times10^{-7}$ \\
        \hline      
        $10^{-3}$ & $1.0\times10^{4}$& $2.1\times10^{-4}$ & $6.8\times10^{-7}$ \\
        \hline   
        $6\times10^{-3}$ & $1.1\times10^{4}$& $2.2\times10^{-4}$ & $7.0\times10^{-7}$ \\
        \hline
    \end{tabular}
    \caption{Limits on the compactification radius (in meters) for $n=1,2,3$ extra dimensions for the three considered red giant profiles, labelled by the zero-age main sequence metallicity.}
    \label{tab:red_giant_limits}
\end{table}

\section{Neutron stars \label{sec:NS}}

Observations of the luminosities of five old neutron stars, whose ages can be inferred from kinematical data and are of order  $10^5$ years, lead to constraints on light BSM particles when compared to predicted cooling curves. The bounds on axions are comparable to those from SN~1987A \cite{Buschmann:2021juv}, while those on new scalars can be up to four orders of magnitude stronger \cite{Fiorillo:2025zzx}. The latter results raise hope that old neutron stars could lead to constraints on KK gravitons that are competitive with supernovae, despite the far fewer KK graviton modes that are accessible due to their much lower temperatures. 
In this section, we show that the resulting limits are actually somewhat weaker than from SN~1987A, although they may still provide a useful complement.

Our analysis follows that for axions of Ref.~\cite{Buschmann:2021juv} and for scalars of Ref.~\cite{Fiorillo:2025zzx} closely, so we provide only the essential inputs; further details and discussion can be found in those references. We use five neutron stars, which have observed ages $t_{\rm obs}$ and photon luminosities at infinity $L_{\rm obs}$ given in Table~\ref{table:NS} (as discussed in detail in Ref.~\cite{Buschmann:2021juv}, the luminosity measurements are more robust than surface temperature measurements that could have alternatively been used).  The ages are inferred from the neutron stars' distances from the assumed position of the supernova events.

\begin{table}[h!]
\centering
\begin{tabular}{|c|c|c|}
\hline
Neutron Star & $t_{\rm obs}$/$10^5\,{\rm yr}$ & $L_{\rm obs}$/$(10^{32}\,{\rm erg}\,{\rm s}^{-1})$   \\ \hline
J1856  & $4.2\pm 0.8$ & $0.65\pm 0.15$ \\
J1308 & $5.5\pm 2.5$ & $3.2\pm 0.6$ \\
J0720 & $8.5\pm 1.5$ & $2.2\pm 1.2$ \\
J1605 & $4.4\pm 0.7$ & $4.0\pm 1.0$ \\
J0659 & $3.5\pm 0.4$ & $2.8\pm 1.4$ \\ \hline
\end{tabular}
\caption{The ages $t_{\rm obs}$ and photon luminosities $L_{\rm obs}$  of the five old neutron stars used in our analysis.}
\label{table:NS}
\end{table}

A cooling neutron star satisfies an energy balance equation 
\beq \label{eq:NSc}
C\frac{dT_c}{dt}=  -L_\gamma^\infty -L_\nu^\infty -L_h^\infty +H,
\eeq
where $L_\gamma^\infty$, $L_\nu^\infty$ and $L_h^\infty$ are the photon, neutrino and graviton luminosities respectively, $T_c$ is the temperature of the core of the neutron star (assumed uniform throughout the core), $C$ is its heat capacity and $H$ encodes possible heating e.g. from the star's magnetic field. In the SM, the cooling at early times  is dominated by emission of neutrinos, which occurs at a rate proportional to $T_c^8$. Subsequently, surface photon emission takes over, usually when the neutron star is about  $10^5$ years old. KK gravitons provide a new cooling channel, resulting in neutron stars that are colder than expected for their ages.

In the absence of superfluidity, the production of KK gravitons in a neutron star's interior is expected to be dominated by bremsstrahlung (processes involving plasmons are strongly Boltzmann suppressed, except for at very early times). The emissivity of neutrino anti-neutrino pairs from neutron-proton and neutron-neutron bremsstrahlung are approximately equal in typical neutron star environments \cite{1979ApJ...232..541F,Bottaro:2024ugp}, and it is reasonable to expect the same is true for KK graviton production by bremsstrahlung. Given that (as discussed subsequently) there are uncertainties of about a factor of two on the neutrino emissivity, for our present purposes it is sufficient to make the conservative approximation of only considering KK graviton production from neutron-neutron bremsstrahlung. 

We evaluate the neutron-neutron bremsstrahlung rate using the soft-radiation approximation for a degenerate neutron environment calculated in Ref.~\cite{Hanhart:2000er}. We further approximate the neutron-neutron scattering cross section, $\sigma_0\simeq 25\,{\rm mb}$, to be independent of energy and scattering angle. The total rate of energy emission per unit time and volume, summed over all KK graviton modes, is then given by
\beq \label{eq:Qgdeg}
\frac{dQ}{dV dt} =  \frac{\pi^{(n-17)/2} \Gamma(n+4)}{270} h_n Y_n \kappa^2  (RT_c)^n \sigma_0 T_c^4 p_F^5 ,
\eeq
where $p_F$ is the neutron Fermi momentum, and 
\begin{align}
    h_n &= \frac{3n^2+18n+19}{(n+5)(n+3)}\frac{1}{\Gamma(\frac{n+3}{2})} , \label{eq:hn}\\
Y_n &= \zeta(n+4)+4\pi^2\frac{\zeta(n+2)}{(n+2)(n+3)}
,
\end{align}
with $\zeta$ the Riemann Zeta function.\footnote{If a more detailed analysis was desired, one could repeat the calculation of Ref.~\cite{Hanhart:2000er} for proton-neutron scattering, with the neutrons in the degenerate limit and protons  non-degenerate.} Although the degenerate and non-relativistic approximations are accurate in neutron stars, we note that Eq.~\eqref{eq:Qgdeg} misses in-medium effects that, e.g., modify the neutron dispersion relation, which might be relevant (these could be interesting to analyse in the future).

We study the cooling of neutron stars using the code NSCool \cite{2016ascl.soft09009P}, which is publicly available and can be straightforwardly modified to include KK graviton emission. This code evolves a spherically symmetric model of a neutron star from early times, soon after formation, until the late times relevant for the observations of the stars that we use. 

There are various nuisance parameters that complicate a comparison between theoretically predicted cooling curves and data. The most important nuisance parameters are the neutron star mass $M_{\rm NS}$, the abundance of light elements in its envelope $\Delta M$, and the equation of state in the core. An additional uncertainty is whether the neutron star's core is superfluid, and if so what its properties are. We adopt the simple approach of assuming that there is no superfluidity and we also neglect any additional heating parametrised by $H$ in Eq.~\eqref{eq:NSc}. These two assumptions are reasonable since they are consistent with current theoretical understanding and lead to a good fit to data in the absence of KK gravitons (as one test of this, we have checked that reversing the sign of the energy emission to KK gravitons, such that they act as a fake energy source, does not substantially improve the fit). If superfluidity was included there would be additional KK graviton production channels involving breaking and formation of Cooper pairs; these however could not lead to constraints unless it was conclusively established that superfluidity was present.

We obtain limits by evaluating the neutron star cooling curves for different values of $m_{\rm KK}$ for 6 neutron star masses linearly spaced between $M_\odot$ and $2M_\odot$, and 8 light element abundances logarithmically spaced between $10^{-20}M_\odot$ and $10^{-6}M_\odot$.  We then construct a likelihood function for the measured data for each of the observed neutron stars, indexed $i$ below, where we assume the errors on the neutron star data are Gaussian,
 \beq
\mathcal{L}^i(L_{\rm obs}^i,t_{\rm obs}^i,\sigma_L^i,\sigma_t^i |m_{\rm KK}, \boldsymbol{\theta}) =\mathcal{N}\left( L(m_{\rm KK}, \boldsymbol{\theta} )-L_{\rm obs}^i, \sigma_L^i \right)\, \mathcal{N}\left( t-t_{\rm obs}^i, \sigma_t^i \right)~,
 \eeq
where $\mathcal{N}(x,\sigma)$ denotes the normal distribution with mean $x$ and standard deviation $\sigma$, and $\boldsymbol{\theta}=(t,M_{\rm NS}, \Delta M)$ are the nuisance parameters. For this purpose, the age of the neutron star $t^i$ at which the predicted luminosity is evaluated is a nuisance parameter. 
Given a value of $m_{\rm KK}$ and set of nuisance parameters for each neutron star (which we collectively denote $\Phi\equiv ({\boldsymbol{\theta}}^i)$) the total likelihood $\mathcal{L}_{\rm tot}(m_{\rm KK},\Phi)$ is the product of the likelihoods for each neutron star.

The best fit value of $m_{\rm KK}$ (or equivalently $R$), which we call $\hat{m}_{\rm KK}$ is obtained by maximising the total likelihood, and we denote the associated values of the nuisance parameters $\hat{\Phi}$. We obtain a lower bound on $m_{\rm KK}$ from the test statistic
\beq
q(m_{\rm KK}) = -2 \ln\left( \frac{\mathcal{L}_{\rm tot}(m_{\rm KK},\Phi_{\rm max}(m_{\rm KK}))}{\mathcal{L}_{\rm tot}(\hat{m}_{\rm KK},\hat{\Phi})} \right),
\eeq
where $\Phi_{\rm max}(m_{\rm KK})$ are the   nuisance parameters that maximise the likelihood for a given $m_{\rm KK}$. 
By Wilks' theorem, the $95\%$ confidence lower bound on $m_{\rm KK}$ is given by $q(m_{\rm KK})=2.7$ (the non-applicability of Wilks' theorem might lead to an order-one error in our limits, which is not a concern for our present purposes \cite{Buschmann:2021juv}).  

We carry out the procedure above for three different choices of equation of state, given the uncertainty on this: the models APR \cite{Akmal:1998cf}, BSk22, and BSk26 \cite{Pearson:2018tkr}. We find that the BSk22 equation of state leads to slightly weaker limits on $m_{\rm KK}$ than the others (by a factor of about $1.5$ for $n=2$ extra dimensions), and it fits the data slightly better in the absence of KK gravitons, so we use this for the limits that we quote.

\begin{table}[t!]
    \centering
    \def\arraystretch{1.3}
    \begin{tabular}{|c|c|c|}
    \hline
          $n=1$  & $n=2$  & $n=3$  \\
         \hline\hline
         $1.5\times10^{3}$& $7.5\times10^{-5}$ & $2.7\times10^{-7}$ \\
         \hline  
    \end{tabular}
    \caption{Limits on the compactification radius (in meters) for $n=1,2,3$ extra dimensions from the five  old neutron stars that we analyse.}
    \label{tab:NS_limits}
\end{table}

\begin{figure}[t!]
    \centering
    \includegraphics[scale=0.445]{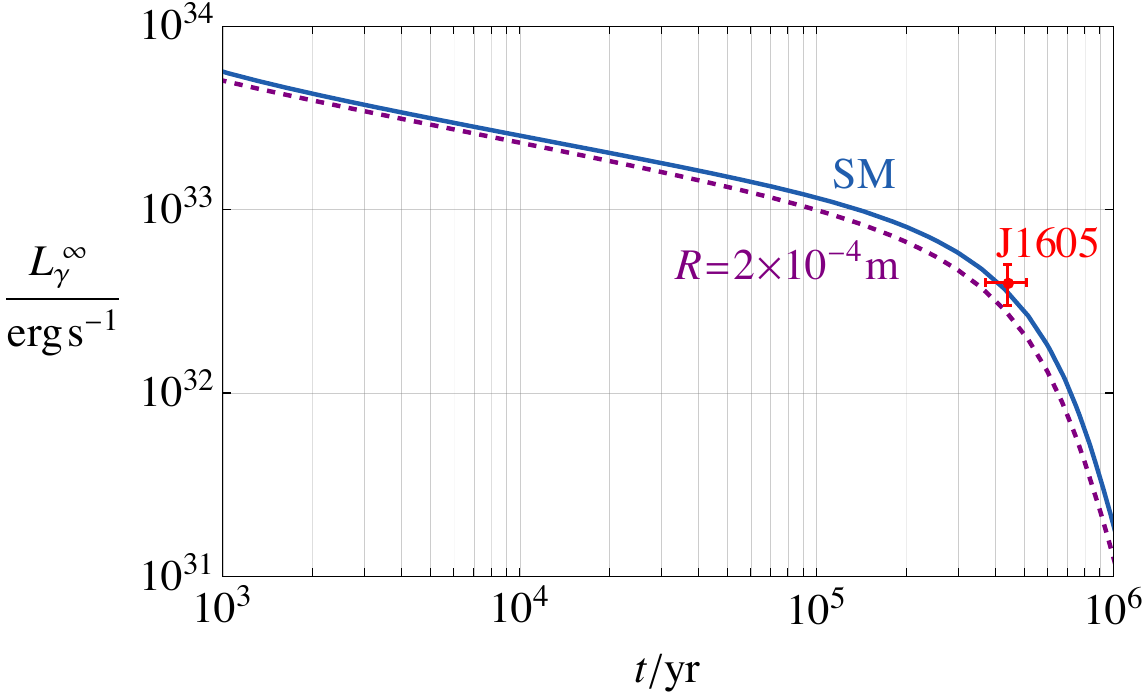}
    \includegraphics[scale=0.445]{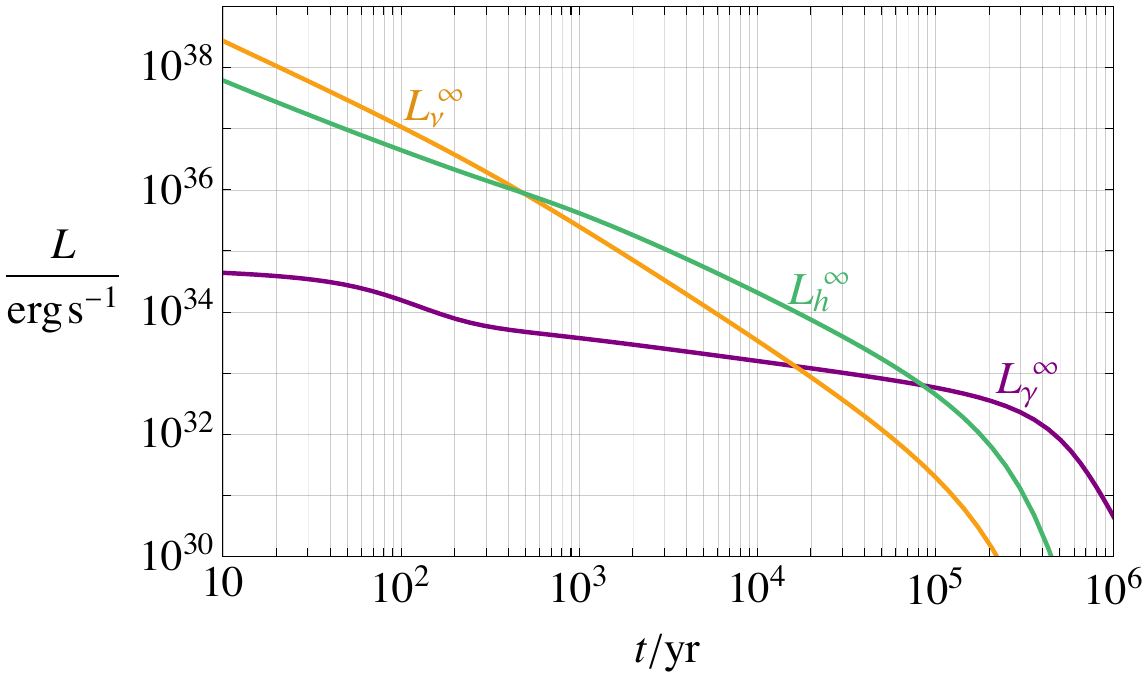}
    \caption{{\it Left:} The observed photon luminosity and age of the neutron star $J1605$ with uncertainties (red), and the best fit cooling curves assuming only Standard Model energy loss (blue, ``SM'') and including also energy loss to KK gravitons with $2$ extra dimensions of radius $2\times 10^{-4}\,{\rm m}$ (dashed purple). Combining the likelihoods over all $5$ neutron stars, the latter theory is excluded. {\it Right:} The energy loss to photons ($L_\gamma^\infty$), neutrinos ($L_\nu^\infty$), and KK gravitons ($L_h^\infty$) in the same large extra dimensional theory as plotted in the left panel. }
    \label{fig:NS}
\end{figure}

The $95\%$ confidence level bounds obtained this way are given in Table~\ref{tab:NS_limits}. As an indication of the impact of KK gravitons on the evolution of a neutron star, in Figure~\ref{fig:NS} we plot the best fit cooling curves for the neutron star J1605 in the absence of KK gravitons and for $n=2$ extra dimensions with $R=2\times 10^{-4}\,{\rm m}$ (i.e. $m_{\rm KK}=10^{-3}\,{\rm eV}$, slightly excluded by our constraint). In both these cases, the best fit curve has a neutron star mass of $M_\odot$ and light element abundance of $10^{-12}M_\odot$. We also show the energy emission to photons, neutrinos and KK gravitons as a function of time. KK graviton emission dominates between approximately $10^4$ and $10^5$ years, around when the dominant SM emission changes from neutrinos to surface photons.

We note that it might be interesting to calculate the production rate in the one-pion-exchange approximation \cite{Cullen:1999hc} with in-medium corrections. This would be fully consistent with the way that neutrino production from bremsstrahlung is treated in NSCool, which builds on the expressions obtained in Ref.~\cite{1979ApJ...232..541F} (Refs.~\cite{Buschmann:2021juv} and \cite{Fiorillo:2025zzx} carry out analyses along these lines for axions and scalars). As mentioned, it would also be useful to include KK graviton production by neutron-proton bremsstrahlung. Given the uncertainties on the equation of state, the fact that the neutrino emission rate likely differs from that implemented in NSCool by factors of about two \cite{Bottaro:2024ugp}, and that the limits we obtain are somewhat weaker than those from SN~1987A, we do not regard these improvements as essential for our present work.

Constraints on the couplings of axions have also been obtained from observations of young neutron stars, such as Cassiopeia A (Cas A) \cite{Leinson:2014ioa,Sedrakian:2015krq,Hamaguchi:2018oqw,Bhat:2019tnz,Leinson:2021ety}, which turn out to be similar to the bounds from old neutron stars. Since the rate of energy emission to axions scales as $T_c^6$, from Eq.~\eqref{eq:Qgdeg} we expect that for $n=1$ and $2$ the limits on KK gravitons from young neutron stars will be comparable to those we have obtained from old stars. For larger $n$ the limits from young neutron stars might be somewhat stronger. However, since the bounds from supernova are much more important for large $n$, we do not consider young neutron stars here (there are presently also uncertainties on the Cas A observational data due to instrumentation issues with the Chandra telescope \cite{Posselt:2018xaf,Buschmann:2021juv}).

\section{SN~1987A}\label{sec: supernova}

During a core-collapse supernova, neutrinos are produced abundantly over a cooling period lasting approximately ten seconds. The high temperatures (up to roughly $50\,{\rm MeV}$) and densities (of order $10^{14}\,{\rm g}\,{\rm cm}^{-3}$) of the protoneutron star during this time make it an ideal environment for the production of low-mass BSM particles. Such particles would lead to additional cooling, shortening the duration of the neutrino cooling era. Despite limited observational data and relatively imprecise theoretical understanding of the dynamics of the explosion, the efficiency of BSM particle production means that supernova constraints are often competitive with other astrophysical bounds, and moreover provide the only cooling limits at higher masses. Even though the relevant values of $m_{\rm KK}$ are small, the high supernova temperatures provide a significant advantage in constraining the large-extra-dimension scenario, because many KK modes are kinematically accessible.

The key observational data comes from SN~1987A, which, to date, is the only core-collapse supernova for which a neutrino burst was detected and can hence be compared to theoretical predictions. Raffelt proposed an analytic constraint that the new energy loss should not exceed $10^{19}\erg\gram^{-1}\sec^{-1}$~\cite{Raffelt:1987yt,Raffelt:1990yz}. This criterion has since been widely applied, including to various theories of new scalars~\cite{Krnjaic:2015mbs,Balaji:2022noj,Hardy:2024gwy}, axions (see \cite{Caputo:2024oqc} and references therein), dark photons (e.g. in \cite{Chang:2016ntp}), as well as KK gravitons~\cite{Cullen:1999hc,Hanhart:2000er,Hanhart:2001fx,Hannestad:2003yd}. 

However, significant uncertainties remain. The explosion mechanism of SN~1987A has not yet been confirmed; although observations are consistent with a remnant neutron star (as expected following core-collapse), other possibilities are not robustly excluded, and in such cases the cooling bound may not apply \cite{Bar:2019ifz}. Even if SN~1987A was a standard core-collapse event, the physics of the explosion is complex and numerical simulations still do not fully capture all processes. As a result, the energy-loss criterion that we use is only reliable to within factors of order-one  (indeed slightly different versions of it are used in the literature). 

We calculate KK graviton production using supernova profiles from three representative spherically symmetric numerical simulations \cite{Rampp:2002bq,Woosley:2002zz,Buras:2005rp} available from the Garching Core-Collapse Supernova Archive~\cite{GarchingArchive}, labelled by their progenitor mass (further discussion of these can be found in Ref.~\cite{Hardy:2024gwy}). We take the profiles immediately prior to when the shock revival is expected to occur and consider energy loss out to the neutrino sphere, at a radius of about $30\,{\rm km}$, as identified in the simulations, imposing the Raffelt criteria in the form above over this volume. The typical core temperatures and densities in the profiles used are consistent with those found in modern three-dimensional simulations during the first few seconds after bounce \cite{Burrows:2024pur}.

We consider three production mechanisms: the well-studied bremsstrahlung channel, as well as the new resonant production and pion-induced processes. 
The resulting upper bounds on the compactification radius, $R$, are summarised in Table~\ref{tab:supernova limits} (the assumptions underlying the two quoted $\pi^-$ process limits are discussed below). In this table, we also show the previous constraints of Refs.~\cite{Hannestad:2003yd} and  \cite{Hanhart:2000er}, which consider bremsstrahlung production using a constant temperature and density supernova model. We see that the limits from the pion process are slightly stronger than those from bremsstrahlung, while those from resonant production are much weaker.

\begin{table}[t!]
    \centering
    \bgroup
    \def\arraystretch{1.3}
 \begin{tabular}{|@{\hspace{0pt}}>{\raggedleft\arraybackslash}p{0.2cm}@{}r|l|l|l|}   
    \hline
         &  & $n=1$  & $n=2$  & $n=3$  \\
         \hline\hline      
         & $17.8 M_\odot$ \qquad \qquad \quad Brem. & $3.0\times10^{2}$& $5.7\times10^{-7}$ & $6.6\times10^{-10}$    \\
        & Res.& $9.6\times10^{4}$  & $2.3\times10^{-5}$  & $1.6\times10^{-8}$   \\
        & $\pi^-$ (a)& $3.6 \times10^{2}$  & $3.8\times10^{-7}$   & $4.1\times10^{-10}$  \\
        & $\pi^-$ (b)& $2.3\times10^{2}$  & $2.7\times10^{-7}$  & $2.6\times10^{-10}$   \\
        \hline
        & $20 M_\odot$ \qquad \qquad \quad Brem.  & $5.3\times 10^2$  & $8.0\times10^{-7}$  & $8.5\times10^{-10}$     \\
        & Res.& 
        $1.3\times 10^{5}$ & $2.7\times 10^{-5}$ & 
        $1.8\times10^{-8}$  \\
        & $\pi^-$ (a)& $7.8\times10^{2}$  & $5.8\times10^{-7}$  & $5.5\times10^{-10}$   \\
        & $\pi^-$ (b)& $5.4\times10^{2}$  & $4.6\times10^{-7}$  & $4.2\times10^{-10}$   \\
        \hline
        &$25 M_\odot$ \qquad \qquad \quad Brem. & $8.3\times10^{1}$  & $2.7\times10^{-7}$  & $3.7\times10^{-10}$    \\
        & Res.& 
        $4.6\times10^{4}$ & 
        $1.6\times10^{-5}$ &
        $1.2\times10^{-8}$\\
        & $\pi^-$ (a)& $5.6\times10^{1}$  &  $1.4\times10^{-7}$  &  $2.0\times10^{-10}$   \\
        & $\pi^-$ (b)& $3.8\times10^{1}$  & $1.2\times10^{-7}$  & $1.8\times10^{-10}$   \\
        \hline
        \hline
         &Hannestad  \& Raffelt \cite{Hannestad:2003yd} & $4.9\times10^{2}$ & $1.0\times10^{-6}$ & $1.1\times10^{-9}$ \\
        \hline
        & Hanhart {\it et al.} \cite{Hanhart:2001fx}  & \cellcolor{gray!10} & $7.1\times10^{-7}$ &  $8.5\times10^{-10}$ \\
         \hline
    \end{tabular}
    \egroup
    \caption{The limits on the compactification radius (in meters) for $n=1,2,3$ extra dimensions for three example supernova models, labelled by their progenitor mass $17.8 M_\odot$, $20M_\odot$ and $25 M_\odot$ (corresponding to protoneutron star masses of $1.59 M_\odot$, $1.50 M_\odot$ and $1.86 M_\odot$ respectively). ``Brem." indicates the limit using only nucleon-nucleon bremsstrahlung emission, and ``Res." indicates the limit using only resonant emission. $\pi^{-}$ (a) and (b) refer to limits considering only the pion process $\pi^-p\rightarrow nh$ with the two sets of approximations described in the main text.}
    \label{tab:supernova limits}
\end{table}

Taking the $17.8M_\odot$ progenitor, which yields the weakest bounds, and combining emission from bremsstrahlung and pions with the more conservative assumptions (a), leads to limits of $m_{\rm KK}\gtrsim 0.6\,{\rm eV}$ for $n=2$ and $m_{\rm KK}\gtrsim 500\,{\rm eV}$ for $n=3$. These are somewhat stronger than those of Refs.\cite{Hanhart:2001fx} and \cite{Hannestad:2003yd}, but, given the substantial uncertainties, the differences are not dramatic. The stronger limits arising from the  $25 M_\odot$ progenitor suggest that improved theoretical understanding of the explosion, and better understanding of the properties of pions in supernovae, might allow stronger limits to be placed. Should we be lucky enough to witness another supernova in our Galaxy or a nearby dwarf galaxy, the resulting high-precision neutrino observations would certainly also be valuable.

\subsection{Bremsstrahlung}

We evaluate the bremsstrahlung production rate following Refs.~\cite{Hanhart:2000ae,Hanhart:2000er} (see also \cite{Hannestad:2003yd})  using the soft-radiation approximation, which is an expansion in the parameter $\chi\equiv \omega m_N/p^2$, where $\omega$ is the energy of the emitted KK graviton and $p^2/m_N$ is the energy of the nucleon-nucleon system (with $p$ the centre of mass momentum and $m_N$ the nucleon mass).\footnote{As shown in \cite{Hanhart:2000er}, the best choice is to take $p^2/m_N$ to be the average of the nucleon energies before and after scattering, since this cancels the next to leading order, $\sim \chi^{-1}$, contribution.} This approximation amounts to considering only the emission from the external nucleon legs, which dominates at small $\chi$ (with rate scaling as $\chi^{-2}$). In general there will be many contributions at order $\chi^0$, which we neglect. We also note that this approach misses multi-body and in-medium correlation effects.

We further assume that the nucleon-nucleon scattering cross section is energy independent and has no dependence on the scattering angle, and that nucleons in the protoneutron star can be approximated as non-degenerate and consist solely of neutrons. 
We expect these approximations to lead to at most corrections of order unity to the rate. The rate of energy emission to KK gravitons by bremsstrahlung is then given by  
\cite{Hanhart:2000er}
\begin{equation}
    \frac{dQ}{dV dt} = \frac{\kappa^2}{18}\sqrt{\frac{T^7}{m_N}}(RT)^n\pi^\frac{n-3}{2}h_n n_n^2 \sigma_0^{nn}\frac{\Gamma(n+1)\Gamma(n+5)}{\Gamma(n+\frac{5}{2})}\frac{n^2+9n+23}{(2n+7)(2n+5)},
\end{equation}
where $h_n$ is defined in Eq.~\eqref{eq:hn}, $n_n$ is the neutron number density,  $\sigma_0^{nn}\approx25\text{mb}$, and we approximate $m_N$ by its vacuum value.

As shown in  \cite{Hanhart:2000er}, at a temperature of $30\,{\rm MeV}$ the expansion parameter $\chi\simeq 3/4$, so in the hottest regions of protoneutron stars (with temperatures around $40\,{\rm MeV}$) the soft approximation is not fully reliable. This is especially the case  because the most massive KK modes are the most numerous for $n>1$. Nevertheless, we regard it as a more reliable and conservative approach than a simple one-pion-exchange calculation, since detailed studies indicate that the latter overestimates the emission rate of axions by up to about an order of magnitude \cite{Chang:2018rso,Carenza:2019pxu} (because, e.g., the corrections from multi-pion exchange are substantial). For the case of KK graviton emission, the one-pion-exchange result also leads to an energy emission rate almost  ten times larger than the soft-approximation result \cite{Hanhart:2000er}. 
A future detailed analysis of the accuracy of the soft approximation would be useful, but is beyond the scope of our present work.

\subsection{Resonant production} 

The rate of KK graviton production by resonant mixing is calculated as described in Sections~\ref{sec:prod} and Appendix~\ref{Appendix: mixing calculation}. For the profiles that we consider, the plasma frequency takes values up to roughly $16\MeV$, which hence sets the maximum KK graviton mass that can be produced by this process. We find that, as for the case of red giants, the longitudinal resonance only dominates for lighter KK modes, and, when summed over the full tower, the transverse resonance is responsible for almost all of the energy production. This effect is more pronounced at larger $n$ since the heavier KK modes have a larger multiplicity.

As can be seen from Table~\ref{tab:supernova limits}, the total rate of energy emission by resonant production is small compared to bremsstrahlung, in contrast to what is found for, e.g., dark photons. We attribute this to the combination of several factors. First, the number of KK modes that are accessible for bremsstrahlung goes as $(RT)^n$, whereas in the resonant process the number scales as $(R \omega_p)^n$, which suppresses the total emission in the regions where $\omega_p\ll T$. In a supernova core, there are regions where $\omega_p\sim T$ due to the large electron chemical potential; however, due to the shape of the photon dispersion, the resonant frequency $\omega_*$ is always greater than $\omega_p$ and hence $T$ by a potentially large factor, unless the graviton mass $m_h\sim\omega_p$. This incurs a large Boltzmann suppression factor $e^{-\omega_*/T}$. Additionally, even focussing on a specific KK mode with $m_h\sim\omega_p$ the expression for the transverse mixing contains a factor of $(1-m_h^2/\omega_p^2)$ which suppresses the production, and similarly the longitudinal production is suppressed (see Eqs.~\eqref{eq:mixrelL} and \eqref{eq:mixrelT} in Appendix~\ref{app:summarymix}). Numerically we find that removing the Boltzmann suppression factor as well as the factor that suppresses the mixing increases the production rate from transverse resonance by a factor of roughly $300$ for $n=2$, explaining most of the observed difference between the resonant and bremsstrahlung limits.\footnote{While the mixing does not have the extra suppression factor for the case of dark photons and scalars, the comparison to KK gravitons is complicated by the fact that we must integrate over a range of particle masses, and in the case of dark photons by the cancellations which occur between the resonant and continuum production, as discussed in Section~\ref{sec:prod}.}

We also note that an interesting related process is the annihilation of two plasmons into a KK graviton. Previous calculations rely on a phenomenological photon mass term~\cite{Das:2008ss}, which violates a Ward identity that we would expect in a consistent thermal field theory calculation, or neglect the plasma mass of the photon~\cite{Barger:1999jf}. The resulting SN~1987A bounds are somewhat weaker than those arising from bremsstrahlung, but are free from nuclear physics uncertainties. It would be beneficial to revisit this by evaluating the $\gamma \gamma h$ vertex consistently within thermal field theory.

\subsection{Production from pions} 

In a cooling protoneutron star, the number density of negatively charged pions $n_{\pi^-}$ is typically a few percent of that of nucleons \cite{Fore:2019wib}. Such a pion number density is present, even though the temperature is well below the pion mass, because there is a substantial isospin asymmetry. This leads to a chemical potential of negative pions  $\mu_{\pi^-}=\mu_n-\mu_p$ (enforced by $n \leftrightarrow p\pi^-$ equilibrium) that is typically well above $50\,{\rm MeV}$ and may even approach $m_\pi$. In contrast, the number density of positive and neutral pions is  negligible. It has recently been shown that the number density of negative pions can be further enhanced by attractive interactions with nucleons in the medium, which reduce the effective pion energy \cite{Fore:2019wib}.

The negative $\pi^-$ abundance in supernovae can lead to axion emission that is comparable to that from bremsstrahlung through  $p\pi^-\rightarrow na$; see \cite{Turner:1991ax,Raffelt:1993ix,Mayle:1987as} for early work of this process and \cite{Carenza:2020cis} for a recent study demonstrating its significance (similar production of dark photons has been considered in Ref.~\cite{Shin:2022ulh}). Here we analyse the analogous channel $p\pi^-\rightarrow n h$. 
It is plausible that such a process might be relatively more important for KK gravitons than axions. This is because the emission spectrum is expected to be harder than bremsstrahlung due to the pion energy, which opens up additional channels in the case of KK gravitons (a recent update to the classic paper \cite{Hannestad:2003yd} highlighted the pion-induced process as potentially relevant for this reason). 

\begin{figure}%
    \centering
    \subfloat[\centering s channel ($M^s$)]
    {{
    \begin{fmfgraph*}(90,60)
    \fmfleft{i2,i1}
    \fmfright{o2,o1}

    \fmfv{label=$\pi^-$}{i1}
    \fmfv{label=$p$}{i2}
    \fmfv{label=$h$}{o1}
    \fmfv{label=$n$}{o2}
    
    \fmf{fermion,label=$p_2$}{i2,v1}
    \fmf{fermion}{v1,v2}
    \fmf{fermion,lab.side=right,label=$p_4$}{v2,o2}
    \fmf{dashes,label=$p_1$}{i1,v1}
    \fmf{dbl_wiggly,lab.side=left,label=$p_3$}{v2,o1}
    \end{fmfgraph*} \vspace{10pt}
    }}%
    \qquad
    \subfloat[\centering u channel ($M^u$)]{{
    \begin{fmfgraph*}(90,60)
    \fmfleft{i2,i1}
    \fmfright{o2,o1}

    \fmfv{label=$\pi^-$}{i1}
    \fmfv{label=$p$}{i2}
    \fmfv{label=$h$}{o1}
    \fmfv{label=$n$}{o2}
    
    \fmf{fermion,label=$p_2$}{i2,v1}
    \fmf{fermion}{v1,v2}
    \fmf{fermion,lab.side=right,label=$p_4$}{v2,o2}

    \fmf{phantom}{i1,v1}
    \fmf{phantom}{v2,o1}
    \fmffreeze
    
    \fmf{dashes,label=$p_1$,lab.side=left}{i1,u1}
    \fmf{dashes}{u1,v2}
    \fmf{dbl_wiggly}{v1,u2}
    \fmf{dbl_wiggly,lab.side=left,label=$p_3$}{u2,o1}
    \end{fmfgraph*} \vspace{10pt}
    }}%
    \qquad
    \subfloat[\centering t channel ($M^t$)]{{
    \begin{fmfgraph*}(90,60)
    \fmfleft{i2,i1}
    \fmfright{o2,o1}

    \fmfv{label=$\pi^-$}{i1}
    \fmfv{label=$p$}{i2}
    \fmfv{label=$h$}{o1}
    \fmfv{label=$n$}{o2}
    
    \fmf{fermion,label=$p_2$,lab.side=left}{i2,v1}
    \fmf{fermion,label=$p_4$,lab.side=left}{v1,o2}
    \fmf{dashes}{v1,v2}
    \fmf{dashes,label=$p_1$}{i1,v2}
    \fmf{dbl_wiggly,label=$p_3$,lab.side=right}{v2,o1}
    \end{fmfgraph*} \vspace{10pt}
    }}%
    \qquad
    \subfloat[\centering contact interaction ($M^c$)]{{
    \begin{fmfgraph*}(90,60)
    \fmfleft{i2,i1}
    \fmfright{o2,o1}

    \fmfv{label=$\pi^-$}{i1}
    \fmfv{label=$p$}{i2}
    \fmfv{label=$h$}{o1}
    \fmfv{label=$n$}{o2}
    
    \fmf{fermion,label=$p_2$,label.side=left}{i2,v1}
    \fmf{fermion,label=$p_4$,label.side=left}{v1,o2}
    \fmf{dashes,label=$p_1$,label.side=left}{i1,v1}
    \fmf{dbl_wiggly,label=$p_3$,label.side=left}{v1,o1}
    \end{fmfgraph*}\vspace{10pt}
    }}%
    \caption{The four Feynman diagrams contributing to the pion process $p \pi^-\rightarrow nh$.}%
    \label{fig:pion diagrams}%
\end{figure}
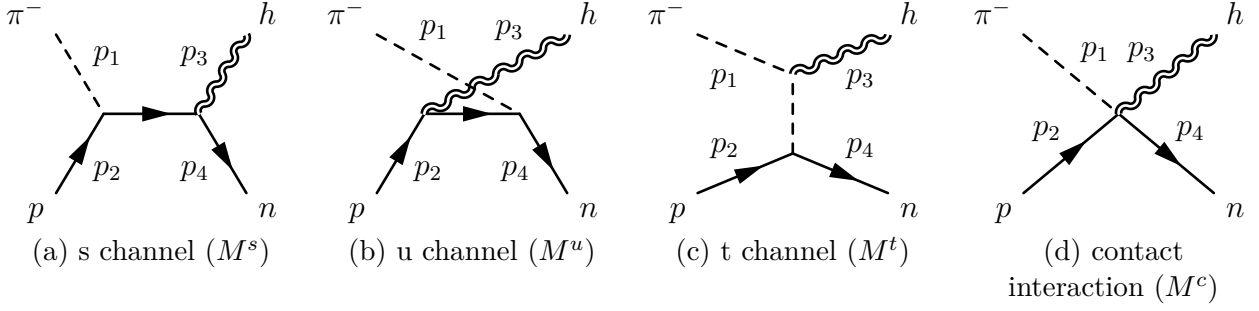

At leading order in chiral perturbation theory, 
the vacuum matrix element for $p\pi^-\rightarrow n h$ is given by the sum of four terms $M_{\alpha\beta} = M^s_{\alpha\beta}+M^u_{\alpha\beta}+M^t_{\alpha\beta}+M^c_{\alpha\beta}$ where $M^c_{\alpha\beta}$ is the 4-point $np\pi^-h$ interaction, corresponding to the Feynman diagrams in Figure~\ref{fig:pion diagrams}. The Feynman rules required to evaluate these are given in  Ref.~\cite{Han:1998sg}, except for the 4-point vertex, which can be derived from the chiral perturbation theory Lagrangian (see Eq.~\eqref{eq:contact} in Appendix~\ref{app:pion}).

\begin{fmffile}{Pion_diagrams}

\end{fmffile}

The expression for the matrix element is given in Appendix~\ref{app:pion}. The total matrix element obeys a Ward identity $p_3^\alpha M_{\alpha\beta}=p_3^\beta M_{\alpha\beta}=0$, which guarantees that terms which diverge as $m_h\rightarrow0$ cancel (this requires that the internal nucleons are not approximated as on-shell). The fermion spin sum can then be performed, as well as sums over graviton polarizations as described in Ref.~\cite{Han:1998sg}. The resulting full expression is long,\footnote{A Mathematica notebook deriving the full expression using the Feyn Calc package~\cite{Mertig:1990an,Shtabovenko:2016sxi,Shtabovenko:2020gxv} is available on request.} and so we take the large $m_N$ limit to give
\begin{align}
    \sum|M|^2 = \frac{\kappa^2 f_{pp}^2m_N^2}{6(x-1)^2}&
    [4(6y-x(6x^2-6xy+y(y+4)+10)) -4z^2(1-x+y)^2
    \\[2ex]\nonumber
    -&z(5x^3+x^2(13-8y)+x(y^2+12y-9)+2y^3-13y^2+12y-9 )] ,
\end{align}
where
\begin{align}
    x = \frac{t}{m_\pi^2},\quad y = \frac{m_h^2}{m_\pi^2}, \quad z = \frac{t-m_\pi^2}{E_\pi^2},
\end{align}
$t = (p_4-p_2)^2$ is the usual Mandelstam variable, with the conventions for the momenta as in Figure~\ref{fig:pion diagrams}, and $E_\pi$ is the pion energy (the unusual variables $x$, $y$ and $z$ are introduced for compactness). 
Numerically,  this large $m_N$ approximation reproduces the full spin summed matrix element up to at worst $\mathcal{O}(1)$ differences.

The production rate of a single KK graviton mode per unit volume and energy $\omega$ is given by
\beq \label{eq:dnhdt}
\frac{d\dot{n}}{d\omega} = \int \frac{d^3 p_2}{(2\pi)^3 2m_N} \frac{d^3 p_1}{(2\pi)^3 2 E_{\pi}} \frac{d^3 p_4}{(2\pi)^3 2m_N} \frac{4\pi \omega^2}{(2\pi)^3 2\omega} (2\pi)^4 \delta^4(p_1+p_2-p_3-p_4) |M|^2 f_p f_\pi (1-f_n)~,
\eeq
where $f_i$ are the Fermi-Dirac and Bose-Einstein distributions as appropriate, and $|M|^2$ is spin summed rather than averaged. Eq.~\eqref{eq:dnhdt} can be partially evaluated (analogously to the calculation in Ref.~\cite{Carenza:2020cis}), such that the total energy emission rate to a single KK mode per unit time and volume  $\frac{dQ}{dVdt}=\int d\omega\,\omega \frac{d\dot{n}}{d\omega}$ is given by
\beq \label{eq:Qh}
\frac{dQ}{dVdt} = \frac{T^{11/2}}{(2\pi)^5 \sqrt{2 m_N}} \left( \frac{z_p z_\pi}{z_n+1}\right) \left(\int_0^\infty dx_p \frac{x_p}{e^{x_p^2}+z_p} \right) \left(\int_{x_{\rm min}}^\infty dx_\pi \frac{x_\pi^2 \epsilon_\pi}{e^{\epsilon_\pi-y_\pi}-z_\pi} \int_{-1}^1 d\cos\theta\,|M|^2 \right),
\eeq
where the fugacities $z_i=\exp((\mu_i-m_i)/T)$; the dimensionless variables  $x_p=p_2/\sqrt{2m_NT}$, $x_\pi=p_1/T$, $y_\pi=m_\pi/T$, $\epsilon_\pi=E_\pi/T$; and $x_{\rm min}$ is set by the minimum pion momentum such that the process is kinematically allowed. $\theta$ denotes the angle between the incoming pion and outgoing graviton in the centre-of-momentum frame, and the kinematics are expanded in small $m_\pi/m_N$, so $t$ is a function only of $E_\pi$ and the masses. A sum over KK modes leads to the total energy emission.

The most robust approach to obtain limits would be to either incorporate KK graviton production directly into a simulation that includes pions, 
or to evaluate Eq.~\eqref{eq:Qh} using a profile  from a simulation that include pions, adopting the same  pion and nucleon equations of state and dispersion relations as used in the simulation.  The matrix element should likewise be evaluated including consistent in-medium corrections.

Unfortunately, this ideal is currently unachievable for several reasons. First, calculating in-medium corrections to the matrix element presents a major challenge. Even for axion emission via bremsstrahlung this is difficult, and KK graviton emission from pions is likely to be even more challenging. Since $\mathcal{O}(1)$ corrections to the leading-order chiral-perturbation-theory result are likely, this limits the accuracy that can realistically be reached.\footnote{Ref.~\cite{Carenza:2020cis} accounts for one such correction for axion emission,  finite nucleon-spin lifetime, by modifying the matrix element with a phenomenological factor \cite{Raffelt:1993ix}, though this effect is relatively minor.} Additionally, even state-of-the-art simulations do not yet include pions.

Given this situation, we adopt an approach expected to provide a reasonable approximation to a full analysis. We use the proton and neutron number-density and temperature profiles from the spherically symmetric simulations discussed above, and we take the pion chemical potential to be $\mu_{\pi^-} = \mu_n - \mu_p$, using the neutron and proton chemical potentials determined in the simulations. 
For simplicity we evaluate Eq.~\eqref{eq:Qh} using the free-particle dispersion relations for the protons and neutrons. We fix the proton and neutron fugacities $z_{n,p}$ to reproduce their number densities as extracted from the simulations with this dispersion relation.

Given the previously discussed medium-induced enhancement of the pion number density, 
we present results using two approaches to the pion abundance, both using $\mu_\pi$ as extracted above, labelled (a) and (b) in Table~\ref{tab:supernova limits}. 
In the first, we use the vacuum pion dispersion relation. 
In the second, following Ref.~\cite{Fore:2019wib}, we take the pion to have a modified dispersion relation 
\beq \label{eq:Epi}
E_{\pi}(p)= \sqrt{p^2+m_\pi^2}+\Sigma_{\pi^-}(p)~,
\eeq
where the self-energy is dominated by interactions with neutrons and is estimated from the one-loop expression
\beq
\Sigma_{\pi^-}(p)=\int \frac{d^3k}{(2\pi)^3} f_{n}(E_{n}(k)) V_{n\pi^-}(p_{\rm cm})~,
\eeq
with $E_n(k)$ the energy of a neutron of momentum $k$, and $p_{\rm cm}$ is the centre-of-mass momentum $p_{\rm cm}= \tilde{m} \left(p^2/m_\pi^2 +k^2/m_N^2- 2{\bf p}.{\bf k}/(m_\pi m_N) \right)^{1/2}$ with $\tilde{m}$ the pion-neutron reduced mass.\footnote{We follow \cite{Fore:2019wib} in using the non-relativistic centre of mass momentum; we have checked that using the relativistic version does not make a significant difference.} The pion–neutron potential is obtained from experimentally measured scattering phase shifts $\delta_{l,\nu}$ as
\beq
V_{n\pi^-}(p_{\rm cm})=-\alpha \sum_{l,\nu} (2l+1)\frac{2\pi \delta_{l,\nu}}{\tilde{m} p_{\rm cm}}~,
\eeq
where $l$ and $\nu$ are, respectively, the angular momentum and nucleon spin projection. 
The factor $\alpha \simeq 0.16$ is phenomenological and is chosen such that, in the limit of small fugacities, the pion number density matches that obtained from a virial expansion. 
Further details can be found in Ref.~\cite{Fore:2019wib}. 
$\Sigma_{\pi^-}$ reduces the pion energy, especially at momenta around $300\,{\rm MeV}$, which enhances the pion abundance for a given $\mu_\pi$.

Additionally, we exclude the central regions of the supernova profiles where the pion chemical potential exceeds $120\,{\rm MeV}$ because the pions are close to condensation in these regions, and both the vacuum dispersion relation and Eq.~\eqref{eq:Epi} are expected to be unreliable.

The first of these two approaches is expected to be conservative, whereas the second is an uncontrolled extrapolation, although it appears to agree well with data in the regimes where comparisons are possible. 
In combination, they are expected to offer a reasonable estimate of the uncertainties and of the range in which the result from a full analysis would likely lie. 
Given the uncertainties, the limits we quote in the abstract and the start of this section are obtained from the first approach (see Appendix F of Ref.~\cite{Fiorillo:2025gnd} for a detailed useful discussion, and for Ref.~\cite{Fore:2023gwv} recent related work).   
As a further cross-check, we have taken the total baryon density and temperature profiles from simulations and computed the neutron, proton, and pion abundances predicted by the virial-inspired model (imposing charge neutrality and beta equilibrium), as in Ref.~\cite{Fore:2019wib}. 
This yields limits that are consistent with the second approach.

The results in Table~\ref{tab:supernova limits} show that the corrected pion dispersion relation leads to a moderate strengthening of the bounds. 
However, the difference between the profiles is more important, primarily due to the higher temperature of the $25M_\odot$ progenitor. 
Consequently, in the future it may be useful to revisit our approach of evaluating the emission only at a single moment rather than through the first few seconds of cooling. 
This would require making use of profiles from three-dimensional simulations, since the spherically symmetric ones we utilise do not successfully produce explosions.

\begin{figure}[t!]
    \centering
    \includegraphics[scale=0.625]{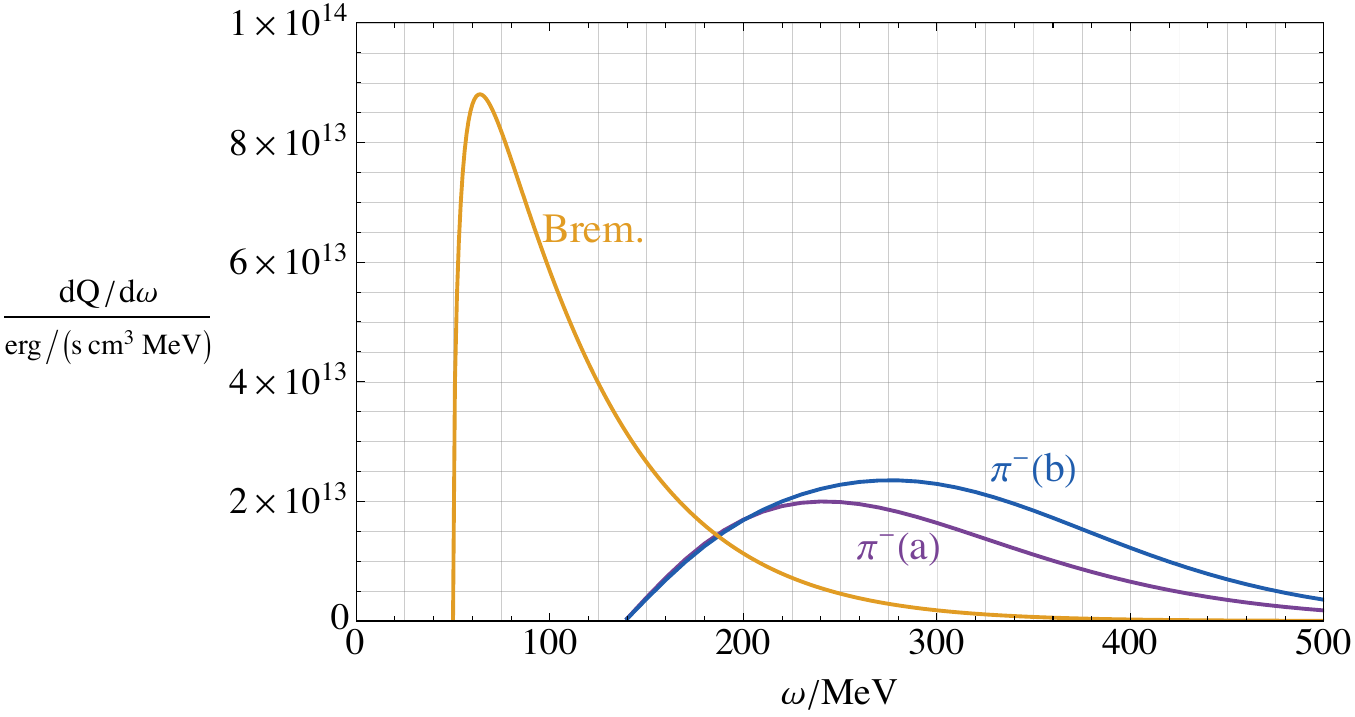} \qquad \qquad \qquad
    \caption{The energy emission spectrum $dQ/(d\omega dVdt)$ to a KK graviton of mass $50\,{\rm MeV}$ (with energy $\omega$) at a typical point in a protoneutron star (see main text). We show the spectra from bremsstrahlung emission (``Brem.'') and the pion process. For the latter, we plot the spectra obtained with the two sets of assumptions described in the main text (labelled ``$\pi^-$(a)'' and ``$\pi^-$(b)'').}
    \label{fig:spec}
\end{figure}

As anticipated, the energy spectrum of KK gravitons produced by the pion process is harder than that from bremsstrahlung.  
To illustrate this, in Figure~\ref{fig:spec} we plot the energy emission spectrum by bremsstrahlung and the pion process for a single KK mode of mass $50\,{\rm MeV}$ at a typical point in the protoneutron star. In particular, we set the temperature to $40\,{\rm MeV}$, the baryon number density to $0.1\,{\rm fm}^{-3}$, the proton fraction to $0.3$, and the pion chemical potential to $80\,{\rm MeV}$. 
The spectrum from the pion process has support only for KK graviton energies greater than the pion mass, because the majority of the incoming pion's energy goes to the KK graviton. 
For the single KK mode shown, the emissivity from bremsstrahlung and the pion process are similar, but the total energy from the latter is greater because its harder spectrum allows more KK modes to be accessed. 
Relatedly, the pion process produces, on average, heavier KK gravitons than bremsstrahlung.

\section{Signals from decays \label{sec:decay}}

In addition to the minimal cooling bounds discussed so far, constraints can arise if some of the KK gravitons produced in an astrophysical object decay to SM particles. 
The decay rate of a KK graviton of mass $m_h$ to approximately massless SM particles is typically
\beq \label{eq:GammaSM}
\Gamma^{\rm SM}= \lambda^2 \frac{m_h^3}{M_{\rm Pl}^2} \simeq (10^{14}\,{\rm yr})^{-1} \lambda^2 \left( \frac{m_h}{{\rm MeV}} \right)^3 ~,
\eeq
where the parameter $\lambda$ encodes SM coupling constants and loop factors, so is not much smaller than one.  
Given Eq.~\eqref{eq:GammaSM}, only the relatively heavy KK gravitons that are produced in supernovae have decay rates fast enough to yield observable signals, while the lighter modes from other types of stars decay far too slowly to be detectable.

Constraints from KK graviton decays to SM particles are strongest when no other decay channels are available. For this to be the case, decays into lighter KK gravitons must be forbidden; whether this is the case is model-dependent. In the limit of an exactly toroidal compactification, such decays are prevented by KK number conservation, the discrete remnant of momentum in the extra dimensions (there is one KK number associated with each extra dimension). In this scenario, constraints arise from various observations associated with supernovae \cite{Hannestad:2003yd}: searches for the products of decays of KK gravitons gravitationally retained by neutron stars (such as RX J185635–3754); searches for the products of decays from the vicinity of the recent supernova remnant Cas A; searches for the products of decays from all past supernovae; and from bounds on the heating of old neutron stars by the products of gravitationally bound KK graviton decays. For $n=2$, these different signals lead to limits on $m_{\rm KK}$ that are a factor of a few to ten stronger than that from the cooling of SN~1987A \cite{Hannestad:2003yd}.

However, as discussed in \cite{Gonzalo:2022jac,Law-Smith:2023czn}, the details of the compactification may induce a small explicit breaking of KK number in the low-energy effective theory. In this case, some of the heavy KK gravitons produced in a supernova would decay to lighter KK gravitons. These lighter particles can still decay to SM particles, but at a much slower rate due to the $m_h^3$ dependence in  Eq.~\eqref{eq:GammaSM}, resulting in weaker constraints. It is therefore interesting to ask whether, with the magnitude of KK number breaking usually assumed in the dark dimension scenario \cite{Gonzalo:2022jac}, constraints from decays into SM particles are relevant.\footnote{
Previously, Ref.~\cite{Anchordoqui:2025nmb} has studied this question. We similarly conclude that limits from decays to SM particles are easily evaded, although our formula for the $n=2$ decay rate differs substantially from theirs.}

We follow Ref.~\cite{Gonzalo:2022jac} in supposing that the details of the compactification allow decays to violate KK number by up to an amount that we call $\delta n$. This is assumed to be much smaller than the KK numbers of typical gravitons produced in a supernova, which is of order  $10\,{\rm MeV}/m_{\rm KK}$. We label the gravitons with KK numbers $(l_1,l_2,\ldots),(m_1,m_2,\ldots),(n_1,n_2,\ldots)$, where the indices correspond to the extra dimensions and  $l_i,m_i,n_i \in \mathbb{Z}$. 
In the absence of KK number violation, the three-point vertex involving KK, is non-zero only if
\beq
l_i= m_i + n_i \quad {\rm for ~all ~}  i~.
\eeq
(in the alternative convention of considering only positive KK numbers the corresponding condition is $l_i= |m_i \pm n_i|$). 
We therefore assume that a KK graviton $(l_i)$ can decay to modes $(m_i),(n_i)$ provided
\beq \label{eq:allowed}
l_i-\delta n< n_i+ m_i<l_i+\delta n \quad {\rm for ~all ~}  i~.
\eeq
Again following \cite{Gonzalo:2022jac}, we suppose that the typical rate for such a decay of a KK graviton of mass $m$ is
\beq \label{eq:Gammad}
\Gamma\simeq \beta^2 \frac{m_h^3}{M_{\rm Pl}^2} \sqrt{\frac{m_{\rm KK}\delta n}{m_h}}~,
\eeq
where $\beta$ is an unknown numerical parameter expected to be $\mathcal{O}(1)$. The factor $\sqrt{m_{\rm KK}\delta n/m_h}$ in Eq.~\eqref{eq:Gammad}, which is the typical velocity of the products of the decays, arises from the phase space since the decays are near threshold.

We focus on the case of two extra dimensions, because for a single extra dimension the astrophysical bounds are much weaker than laboratory constraints, even if KK gravitons decay exclusively to SM particles \cite{Hannestad:2003yd}. We assume that equal KK number violation is possible in all internal directions, as in Eq.~\eqref{eq:allowed}.\footnote{In the case of two extra dimensions, but with KK number violation only in one of the directions, the decay rate is a factor of $\delta n$ smaller than Eq.~\eqref{eq:Gammatot}. However, our broad conclusions  hold also in this case.} In Appendix~\ref{app:decay}, we show that for $n=2$ the number of decay channels for a KK mode of mass $m$ is approximately
\beq
\left( \frac{m_h}{m_{\rm KK}} \right)^{3/2} \delta n^{5/2}~.
\eeq
Consequently, the total decay rate to KK gravitons is 
\beq \label{eq:Gammatot}
\Gamma^{\rm tot}(m) \simeq \beta^2 \frac{m_h^4 \delta n^3}{M_{\rm Pl}^2 m_{\rm KK}} ~,
\eeq
which corresponds to a typical lifetime of
\beq \label{eq:tauSM}
\frac{1}{\Gamma^{\rm tot}} \simeq 0.1\,{\rm yr} \left(\frac{10}{\beta}\right)^2\left(\frac{10}{\delta n} \right)^{3} \left(\frac{10\,{\rm MeV}}{m_h} \right)^{4} \left(\frac{m_{\rm KK}}{{\rm eV}} \right)~.
\eeq

To calculate precise limits and predictions, one must  account for the distribution of KK graviton energies and masses, as well as time-dilation, but for our present purposes a rougher analysis is sufficient. For $n\geq 2$ most of the KK gravitons produced have masses not too much smaller than their energies, because the number of KK modes increases with mass, so we neglect the effect of time-dilation. We also approximate the mass distribution of KK gravitons from a supernova as a delta-function peaked at $m_h(t)$. Moreover, we assume that each KK graviton decays into two products of equal mass, each of mass $m_h/2$ (as shown in Appendix~\ref{app:decay}, this is a reasonable approximation for the majority of decays). The typical graviton mass as a function of time then satisfies 
\beq
\frac{dm_h(t)}{dt}\simeq -\Gamma^{\rm tot}(m_h) \frac{m_h(t)}{2}~.
\eeq
Hence, given an initial population of KK gravitons of mass $m_0$, we have
\beq \label{eq:mt}
m_h(t) \sim m_0 \left(1 + \Gamma^{\rm tot}(m_0)t/2 \right)^{-1/4}~,
\eeq
and the corresponding comoving number density of KK gravitons is $n(t) \simeq n_0 m_0/m_h(t)$, where $n_0$ is the initial number density.

Suppose that an observational signal arises from energy injection into the SM at a time $t_{\rm obs}$ after a supernova. 
Then the corresponding rate of energy transfer to the SM, $\dot{E}_{\rm SM}$, is
\beq
\begin{aligned}
\dot{E}_{\rm SM}(t_{\rm obs}) &\simeq n(t_{\rm obs}) m_h(t_{\rm obs}) \frac{\lambda^2 m_h(t_{\rm obs})^3}{M_{\rm Pl}^2}~.
\end{aligned}
\eeq
Relative to the rate in the absence of KK number violation, $\dot{E}_0$, we have\footnote{For some signals, such as the background from all galactic supernovae or heating of old neutron stars, the relevant quantity is the total energy injected into the SM up to $t_{\rm obs}$. This too is strongly suppressed relative to the case of purely SM decays for $\Gamma^{\rm tot}(m_0)t_{\rm obs}\gg 1$.} 
\beq  \label{eq:reld}
\begin{aligned}
\frac{\dot{E}_{\rm SM}(t_{\rm obs})}{\dot{E}_0} &\simeq \left(1 + \Gamma^{\rm tot}(m_0)t_{\rm obs}/2 \right)^{-3/4}~.
\end{aligned}
\eeq

The typical timescale relevant to the heating of old neutron stars is roughly $10^5$ years, similar to that for direct searches for the decay products of gravitationally bound KK gravitons. The corresponding timescale for all past supernovae is even longer. Since these observations provide limits on $m_{\rm KK}$ that are not dramatically stronger than those from SN~1987A cooling even in the case of decays only to the SM, from Eqs.~\eqref{eq:tauSM} and \eqref{eq:reld} it follows that for $\delta n$ of at least a few these constraints are much weaker than cooling bounds.

An interesting exception is the signal from Cas A, which has the shortest relevant timescale $t_{\rm obs}\simeq 300\,{\rm yr}$. Although previous analyses give constraints only about as strong as those from SN~1987A cooling, this could be improved by future observations. From Eqs.~\eqref{eq:tauSM} and \eqref{eq:reld}, for $\delta n\simeq 10$ decays to other KK gravitons render this signal irrelevant. However, if $\delta n$ is only slightly larger than $1$ it could still be important.

Additionally, previous limits considered only KK gravitons produced by bremsstrahlung. However, the pion production process is likely to be particularly important for these searches, since (as shown in Figure~\ref{fig:spec}) it yields a substantially harder energy spectrum than bremsstrahlung. Consequently, the typical KK graviton mass is larger by roughly a factor of three, leading to decay rates (both intra-tower and to SM particles) that are 
more than an order of magnitude faster. The resulting SM decay products will also be more energetic, and therefore easier to detect above the background. Together, these effects could significantly enhance the sensitivity of future searches.

Meanwhile, for $n>2$ the number of available decay channels for a given KK graviton mass is even larger, and the signals from decays to the SM are correspondingly weaker than those from cooling for $\delta n$ non-zero. 

We also note that an alternative realisation of KK number violation has been proposed and studied in detail in Ref.~\cite{Mohapatra:2003ah}, in which the KK modes of a bulk scalar acquire expectation values. This leads to a different pattern of allowed decays than we have assumed. It might be interesting to explore the implications of this setup for the dark dimension scenario.

\section{Discussion \label{sec:conc}}

In this paper we have revisited astrophysical cooling constraints on KK gravitons, motivated by recent interest in the dark dimension scenario. After analysing a wide range of astrophysical systems and considering two novel production mechanisms, we conclude that the bound from SN~1987A remains the most stringent. In this environment, even with conservative assumptions, the rate of production of KK gravitons from pions exceeds that from bremsstrahlung by a factor of a few, provided the soft approximation is accurate for the latter. As can be seen from Table~\ref{tab:supernova limits}, uncertainty in the protoneutron star profile of SN~1987A is a critical factor. Because of several partial cancellations, resonant production is relatively unimportant in supernovae, and it leads to limits from red giants that are a few orders of magnitude weaker than those from SN~1987A.\footnote{In passing, we note that the calculation of the in-medium mixing, and the cancellation of unphysical $1/m_{h}$ divergences, is somewhat delicate, see Appendix~\ref{Appendix: mixing calculation}. However, this is guaranteed from a Ward identity (useful related work can be found in Refs.~\cite{Gill:2023kyz,Chivukula:2023sua,deGiorgi:2023mdy}).} Cold neutron stars  also lead to constraints that are somewhat weaker than SN~1987A for $2$ and $3$ extra dimensions.

There are several ways that the robustness of the SN~1987A limit could be improved, although these are likely to be technically involved. The calculation of the bremsstrahlung rate using the soft approximation could be refined by analysing the impact of multi-body interactions or evaluating corrections that are formally higher order in the expansion but might be numerically significant. The rate could also be computed by evaluating the matrix elements directly, modifying the one-pion-exchange approximation with corrections or more sophisticated scattering potentials, building on work done in the context of axions, e.g., in Ref.~\cite{Carenza:2019pxu}. It would be especially useful to compute corrections to the leading-order chiral perturbation theory prediction for the pion process. On the simulation side, given the importance of pions for both KK graviton and axion limits, the inclusion of these in simulations would be valuable. Ultimately, full three-dimensional simulations including new sources of energy loss would be ideal, as in e.g. \cite{Betranhandy:2022bvr}. However, significant progress is required even in the purely SM case, and the relevant BSM microphysics must be reliably determined before such extensions are worthwhile. The production of KK gravitons might also lead to observable effects in the evolution of supernova, beyond conventional cooling \cite{Caputo:2025aac}.

We have also shown that if decays within the KK graviton tower proceed at the rate typically assumed in the dark dimension scenario, searches for SM decay products lead to limits that are weaker than those from cooling bounds (although they remain significant if KK number violation is highly suppressed). Among the various proposed signals, the most promising are searches for SM particles from around very young neutron stars, where the population of heavy KK gravitons has had the least time to deplete. This is encouraging, since future observations of Cas A might provide sensitivity to previously unconstrained theories (the remnant of SN~1987A could also be observed, although it is further from Earth). As discussed, the pion production process could be especially important in this context, and a detailed study would be interesting.

For one extra dimension, cooling constraints from old neutron stars are comparable to those from SN~1987A, but both are substantially weaker than laboratory bounds. 
In this case, the best prospects for stronger future limits likely come from improved experiments searching for deviations from Newtonian gravity (on the other hand, one can hope for a discovery in specific non-minimal models that yield stronger signals, e.g. if there are neutrinos \cite{Anchordoqui:2024xvl,Elacmaz:2025ihm, Eller:2025lsh} or axions \cite{Mario} propagating in the 5 dimensional bulk). It is interesting to note that improved short-range gravity experiments could strengthen the lower bound on $m_{\rm KK}$ by about an order of magnitude, which would probe part of the dark dimension parameter space (see, e.g., \cite{Manley:2024svv}). Meanwhile, for two extra dimensions, a possible future $100\,{\rm TeV}$ particle collider might be sensitive to theories currently allowed by SN~1987A constraints.

A notable omission from our present work is constraints from white dwarfs. Such systems lead to relatively strong bounds on axions, both from the luminosity function \cite{Raffelt:1985nj,Blinnikov:1994eoa,MillerBertolami:2014rka,Isern_2018} and from the observed periods of pulsating white dwarfs \cite{Corsico:2012sh,Corsico:2016okh,Battich:2016htm,Isern:2010wz}. Limits on KK gravitons have been obtained from pulsating white dwarfs in the past \cite{Biesiada:2001iy}, and it would be interesting to revisit these, although the previous limits were weaker than those from the cooling of SN~1987A. The resonant production process that we have considered might play a role, especially at high luminosities where neutrino cooling proceeds mainly due to plasmon decay. A reliable limit from the luminosity function would require implementing KK graviton emission in a white dwarf evolution model to capture the effects of feedback \cite{MillerBertolami:2014rka}, as well as careful analysis of the microphysics (e.g. ionic effects on bremsstrahlung emission \cite{Nakagawa:1987pga}). Given that observational surveys are providing new data on the luminosity function \cite{2015yCat..74464078K}, this would be worthwhile. Pulsating white dwarfs are especially promising for KK gravitons since they can be relatively hot, and again new data is expected. It has also recently been shown that magnetic fields, such as those around magnetic white dwarfs, can affect resonant conversion of axions \cite{Caputo:2020quz,Caputo:2021kcv,Brahma:2025vdr}, and it might be interesting to investigate the consequences of this for KK gravitons.

Finally, although not the focus of our present work,  several aspects of the cosmology of KK gravitons merit a brief mention. 
If KK gravitons produced from the SM thermal bath in the early Universe are effectively stable, then for $n=2$, a low reheating temperature of order $5\,{\rm MeV}$ is required to avoid overclosing the Universe, close to the minimum that is consistent with Big Bang Nucleosynthesis. 
If there are decays within the KK graviton tower with small $\delta n$, the decay products are approximately non-relativistic, and this bound on $T_{\rm RH}$ is essentially unchanged. For larger $\delta n$, the decay products are too warm to constitute all of dark matter. However, if $\delta n$ is sufficiently large, the decays might be fast enough that the KK graviton population remains relativistic after production, causing its energy density to redshift like radiation. It would be interesting to explore whether such scenarios can lead to viable cosmological histories with the KK gravitons as a warm, subdominant,  component of dark matter. This population may, for example, contribute to $\Delta N_{\rm eff}$ at a level detectable with the forthcoming CMB-S4 experiment. It would also be interesting to study the signals of decays of dark matter KK gravitons to the SM in the $n=2$ case, analogous to the $n=1$ analysis of Ref.~\cite{Law-Smith:2023czn}.

\section*{Acknowledgements}
We thank Georges Obied and Mario Reig for useful discussions. EH acknowledges the UK Science and Technology Facilities Council for support through the Quantum Sensors for the Hidden Sector collaboration under the grant ST/T006145/1 and UK Research and Innovation Future Leader Fellowship MR/V024566/1. AS is funded by the UK Research and Innovation grant MR/V024566/1. HS is supported by The Science and Technology Facilities Council under the grant ST/X508664/1.
We acknowledge the COST Action COSMIC WISPers CA21106, supported by COST (European Cooperation in Science and Technology).
For the purpose of open access, the author has applied a CC BY public copyright license to any Author Accepted Manuscript (AAM) version arising from this submission.

\appendix

\section{Photon--KK graviton mixing}\label{Appendix: mixing calculation}
The photon--KK graviton mixing receives contributions from each species of charged fermion in the plasma $\Pi^{\mu,\alpha\beta}_{\gamma h}(Q) = \sum_f\Pi^{\mu,\alpha\beta}_{\gamma h, f}(Q)$. Using the Feynman rules in Ref.~\cite{Han:1998sg}, and performing a calculation in the framework of real-time thermal field theory as described in Ref.~\cite{Hardy:2024gwy}, we find that the mixing due to a fermion of charge $q_f$ with energy-momentum relation $E_p^2 = p^2+m_f^2$ is given by
\begin{align}\label{eqn: Full mixing}
    \Pi^{\mu,\alpha\beta}_{\gamma h, f}(Q)=q_f e \kappa \int \frac{d^3 p}{(2\pi)^3}\frac{1}{2E_p}\left(f_e(E_p)-f_{\bar{e}}(E_p) \right)\left(A^{\mu\alpha\beta}+\frac{B^{\mu\alpha\beta}+C^{\mu\alpha\beta}}{4(P\cdot Q)^2-Q^4}\right),
\end{align}
where
\begin{equation}
    A^{\mu\alpha\beta} =\frac{2 Q^\mu}{m_A^2 \xi}(P^\alpha Q^\beta + P^\beta Q^\alpha-g^{\alpha \beta}(P\cdot Q))-(P^\alpha g^{\mu\beta}+P^\beta g^{\mu\alpha}),
\end{equation}
\begin{align}
    B^{\mu\alpha\beta}  =  2&P^\mu\left[g^{\alpha \beta}\left(4(P\cdot Q)^2-Q^4\right)-4(P\cdot Q)\left(P^\alpha Q^\beta + P^\beta Q^\alpha\right)+Q^2 Q^\alpha Q^\beta\right]\\
    +&Q^\mu\left[Q^2\left(P^\alpha Q^\beta + P^\beta Q^\alpha\right)-8(P\cdot Q)P^\alpha P^\beta\right]\\
    -&Q^2 (P\cdot Q)\left(Q^\alpha g^{\mu \beta} + Q^\beta g^{\mu \alpha}\right),
\end{align}
and
\begin{equation}
    C^{\mu\alpha\beta} = 4(P\cdot Q)^2\left(P^\alpha g^{\mu\beta}+P^\beta g^{\mu\alpha}\right)+8 Q^2 P^\mu P^\alpha P^\beta.
\end{equation}
Here we have introduced a photon mass $m_A$ which is necessary to regulate the $t$ channel diagram, and $\xi$ is the usual gauge fixing parameter appearing in the photon's propagator. The photon's Lorentz index is $\mu$, and the KK graviton's Lorentz indices are $\alpha\beta$. The $A^{\mu \alpha \beta}$ term comes from the $t$ channel and four-point contact diagrams shown in Table~\ref{tab:SelfEnergyDiagramTable}, while the $B^{\mu \alpha \beta}$ and $C^{\mu \alpha \beta}$ terms both come from the s-channel and u-channel diagrams.

Eq.~\eqref{eqn: Full mixing} will be contracted with physical photon and KK graviton polarization vectors, which have the properties
\begin{equation}
    \epsilon^\gamma_\mu Q^\mu = 0,\qquad  \epsilon^h_{\alpha\beta}Q^\alpha=\epsilon^h_{\alpha\beta}Q^\beta=0,\qquad \epsilon^h_{\alpha\beta}g^{\alpha\beta}=0.
\end{equation}
Therefore it is easy to see that the $B^{\mu \alpha \beta}$ term will not contribute to the mixing. Furthermore, the $A^{\mu \alpha \beta}$ term reduces to $-(P^\alpha g^{\mu\beta}+P^\beta g^{\mu\alpha})$ so $\xi$ will not enter in any physical quantities, and the remaining terms are proportional to
\begin{equation}\label{eqn: vanishing P integral}
    q_f\int\frac{d^3p}{(2\pi)^3}\frac{P^\alpha}{2E_p}\left(f_e(E_p)-f_{\bar{e}}(E_p) \right).
\end{equation}
By rotational symmetry, only the $\alpha=0$ component can survive, however $P^0 = E_p$ which cancels the term in the denominator and so the integral is simply proportional to $q_f (n_f-n_{\bar{f}})$. When we sum over the different fermion species, this gives the total charge density in the plasma, which must be zero as we have assumed that the plasma is in equilibrium. This means that $A^{\mu\alpha\beta}$ also does not contribute. Only the s-channel and u-channel contribute to the mixing, and only the $C^{\mu\alpha\beta}$ term remains. We can see that this is due to the presence of a tadpole in the thermal loop diagrams corresponding to the other channels. 

Before we move on to perform the integration over momenta, we note some features of $\Pi^{\mu,\alpha\beta}_{\gamma h}(Q)$. Firstly we have the following Ward identities, also identified in Ref.~\cite{Gill:2023kyz}, corresponding to the electromagnetic U(1) symmetry and gravitational diffeomorphisms\footnote{Additionally, due to the charge conjugation properties of the graviton and the photon, the mixing vanishes if $f_e(E_p)=f_{\bar{e}}(E_p)$ since the plasma is invariant under C-symmetry meaning we may apply Furry's theorem.}
\begin{equation}
    Q_\mu\epsilon^h_{\alpha\beta}\Pi^{\mu,\alpha\beta}_{\gamma h}(Q)=0
\end{equation}
\begin{equation}
    \epsilon^\gamma_\mu Q_\alpha Q_\beta \Pi^{\mu,\alpha\beta}_{\gamma h}(Q)=0.
\end{equation} 
Naively, contracting every index with the external momentum appears to give a non-zero value; however, once again after summing over all the fermion species in the plasma, charge neutrality enforces the identity
\begin{equation}
    Q_\mu Q_\alpha Q_\beta  \Pi^{\mu,\alpha\beta}_{\gamma h}(Q) = 0.
\end{equation}
These identities ensure that contributions at order $1/m_h$ vanish. While we expect this to hold to higher orders in perturbation theory, we have only checked the leading order case.
It can also be checked that mixing only occurs between modes with the same helicity. For example the longitudinal polarization state of the photon does not mix with the helicity-1 states of the graviton, $\epsilon_\mu^0\epsilon_{\alpha\beta}^{\pm}\Pi^{\mu,\alpha\beta}_{\gamma h}=0$.
To evaluate the integral, we assume $4(P\cdot Q)^2-Q^4\approx 4(P\cdot Q)^2$, originally argued for in Ref.~\cite{Braaten:1993jw}. This means that the denominator cancels in the first term of $C^{\mu\alpha\beta}$, leaving something of the same form as Eq.~\eqref{eqn: vanishing P integral}, which vanishes. The only part which contributes to the final mixing is therefore
\begin{align}
    \Pi^{\mu,\alpha\beta}_{\gamma h, f}(Q)_{\text{reduced}}=2 Q^2 q_f e \kappa \int \frac{d^3 p}{(2\pi)^3}\frac{1}{2E_p}\left(f_e(E_p)-f_{\bar{e}}(E_p) \right)\frac{P^\mu P^\alpha P^\beta}{(P\cdot Q)^2}.
\end{align}
 After contracting with appropriate polarization vectors, the angular integration can be performed explicitly, yielding expressions of the form
\begin{equation}
    \Pi^{L,T}_{\gamma h, f}(Q)=q_f e \kappa \int_0^{\infty} p^2\left(f_e(E_p)-f_{\bar{e}}(E_p) \right) H^{L,T}(E_p,p,\omega,k) dp,
\end{equation}
where $Q = (\omega,\boldsymbol{k})$ and $L,T$ denotes the mixing for the longitudinal and transverse modes respectively. We are free to shift our definition of $H^{L,T}$ by any constant $H^{L,T}\rightarrow H^{L,T}+f(\omega,k)$ since this will change the mixing by an amount proportional to the charge density which cancels out upon summing over species. We find that a convenient choice is to subtract off the zero-momentum contribution and define
\begin{equation}
    \bar{H}^{L,T}(E_p,p,\omega,k)=H^{L,T}(E_p,p,\omega,k)-H^{L,T}(m_f,0,\omega,k).
\end{equation}
This both removes the unphysical $1/m_h$ divergence in the contribution from each individual species, and makes it easier to keep track of which species are important, since now in the non-relativistic limit $\bar{H}^{L,T}\propto p^2/m_f^2+\mathcal{O}(p^4/m_f^4)$, meaning that the non-relativistic contribution is proportional to the pressure, which is suppressed by $T/m_f$. 
Since the $1/m_h$ terms only cancel out when overall charge neutrality is assumed, it is interesting to ask if such terms can exist out of equilibrium. These would lead to a large enhancement of the KK graviton production rate, and therefore to interesting observational signatures. We expect that the answer is no, and that the cancellation is enforced by Ward identities which rely only on the gauge symmetries of the theory. For a detailed discussion of the Ward identities and associated power counting see Ref.~\cite{Chivukula:2023qrt}.

\subsection{Summary of mixing calculation} \label{app:summarymix}
We now present the final results of the calculation of the photon--KK graviton mixing $\Pi_{\gamma h,f}^{L,T}$ in various regimes. If the fermion, $f$, is non-degenerate and non-relativistic so that $p/m_f\ll1$ for typical momenta, we find
\begin{equation}
    \Pi_{\gamma h,f}^{L}(\omega) = -\frac{1}{\sqrt{6}}q_f e \kappa \frac{n_f T}{m_f}\frac{m_h}{\omega}v(3v^2-2),
\end{equation}
\begin{equation}
    \Pi_{\gamma h,f}^{T} = -\frac{1}{\sqrt{2}}q_f e \kappa \frac{n_f T}{m_f}\frac{m_h}{\omega}v,
\end{equation}
where we have simplified the result using $\omega^2-k^2=m_h^2$, and introduced the graviton velocity $k = v \omega$. For the case where the fermions are ultra-degenerate, so that the distribution function can be approximated as a step function at the Fermi momentum $p_F$, the mixing is given by
\begin{equation}
    \Pi_{\gamma h,f}^{L} = \frac{q_f e \kappa}{6\sqrt{6}\pi^2}p_F^3\frac{m_h}{v \omega}\left((3-2v^2)H(v^2 v_F^2)-v^2v_F^2(1+H(v^2 v_F^2)\right),
\end{equation}
\begin{equation}
    \Pi_{\gamma h,f}^{T} =-\frac{q_f e \kappa}{3\sqrt{2}\pi^2}p_F^3\frac{m_h}{v \omega}\frac{G(v^2 v_F^2)}{2},
\end{equation}
where $H(x)$ and $G(x)$ are functions which commonly appear in photon self-energy calculations \cite{Raffelt:1996wa,Braaten:1993jw} and are defined as
\begin{equation}
G(x) = \frac{3}{x}\left( 1-\frac{2x}{3} -\frac{1-x}{2\sqrt{x}}\log\left({\frac{1+\sqrt{x}}{1-\sqrt{x}}}\right)\right), \qquad H(x) = \frac{G(x)-x}{x-1}.
\end{equation}
The Fermi velocity $v_F$ is defined as $p_F = v_F E_F$. Provided that the species in question also gives the dominant contribution to the photon's self energy, the logarithms appearing in $H(x)$ and $G(x)$ can be eliminated by using the resonance condition $\Pi^{L/T}_{\gamma \gamma}(\omega_*,k_*)=\omega_*^2-k_*^2=m_h^2$ to give\footnote{The typical velocity $v_0$ appearing in $\Pi_{\gamma\gamma}$ is taken to be $v_F$.}
\begin{equation}\label{eqn: L deg mixing simplified}
    \Pi_{\gamma h,f}^{L} = \frac{q_f e \kappa}{6 \sqrt{6} \pi^2}p_F^3
\frac{\omega_* m_h}{v_* \omega_p^2}\left((3-2v_*^2)\left(1-\frac{\omega_p^2}{\omega_*^2}\right)-v_*^2 v_F^2\right),
\end{equation}
 \begin{equation}\label{eqn: T deg mixing simplified}
    \Pi_{\gamma h,f}^{T} = \frac{q_f e \kappa}{3 \sqrt{2} \pi^2} p_F^3\frac{m_h}{v_* \omega_*}\left(1-\frac{m_h^2}{\omega_p^2}\right),
\end{equation}
where $v_*=k_*/\omega_*$. In all cases studied here, the electrons dominate both the mixing and the photon self-energy. However, one may imagine a situation in which one species is dominant in the photon self-energy, and a different species is dominant in the mixing. In this case Eqs.~\eqref{eqn: L deg mixing simplified} and \eqref{eqn: T deg mixing simplified} would not apply. Notice that if the dominant species in the plasma is non-relativistic, such that $v_F\approx0$, the longitudinal resonance satisfies $\omega_* \approx \omega_p$ and the transverse resonance satisfies $m_h \approx \omega_p$. In this limit both expressions vanish, as expected to match the non-relativistic case at lowest order in $T/m_f$.
Making no approximations about degeneracy, but taking the fermions to be ultra-relativistic, we find after eliminating the logarithms with the resonance condition (now taking $v_0=1$)
\begin{equation} \label{eq:mixrelL}
    \Pi_{\gamma h,f}^{L} = \frac{q_f e \kappa n_f}{2 \sqrt{6}}
\frac{\omega_* m_h}{v_* \omega_p^2}\left((3-2v_*^2)\left(1-\frac{\omega_p^2}{\omega_*^2}\right)-v_*^2 \right),\end{equation}
 \begin{equation} \label{eq:mixrelT}
    \Pi_{\gamma h,f}^{T} = \frac{q_f e \kappa n_f}{\sqrt{2}} \frac{m_h}{v_* \omega_*}\left(1-\frac{m_h^2}{\omega_p^2}\right),
\end{equation}
which agrees with taking the relativistic limit of the ultra-degenerate expression. We again assume that the fermion, $f$, is the dominant contributor to the photon's self energy.

If all particles are ultra-relativistic, then all contributions to the mixing are proportional to $q_fn_f$ so the leading order result shown here once again cancels out due to charge neutrality. Higher order terms are then suppressed by a factor of $m_f^2/T^2$, which is small in the ultra-relativistic limit.

\section{Red giant profiles}\label{app:RG}

\begin{figure}[t!]
    \centering
        \subfloat
    {{\includegraphics[width=0.55\textwidth]{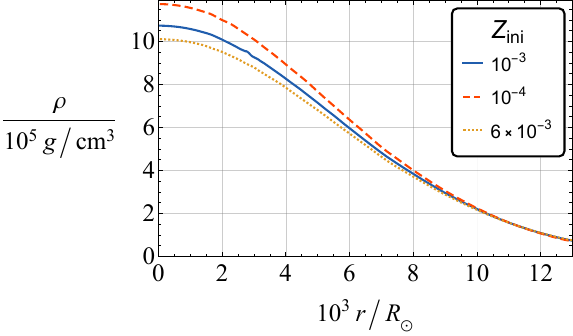} \qquad \qquad \qquad  }}  \\
    \vspace{0.5cm}
    \subfloat
    {{\includegraphics[width=0.475\textwidth]{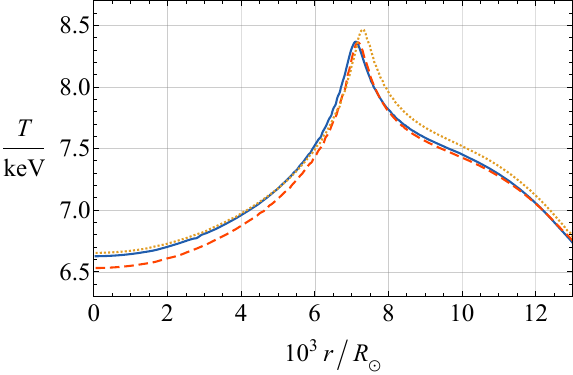} }} \quad
    \subfloat
    {{\includegraphics[width=0.475\textwidth]{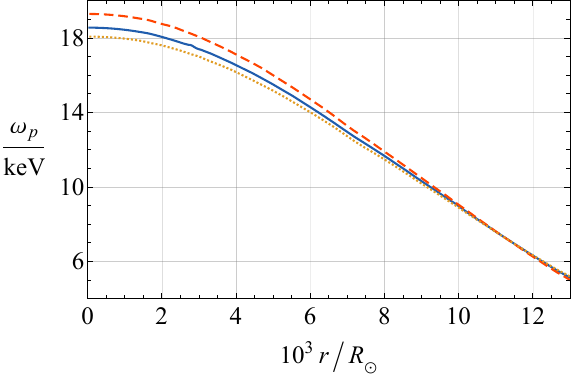} }}\\

    \caption{The density $\rho$,  temperature $T$, and plasma frequency $\omega_p$ as a function of radial coordinate $r$ in the red giant cores for the 3 different metallicities, $Z_{\rm ini}$, that we use.} 
    \label{Fig: RG profiles} 
\end{figure}

The profiles for the cores of stars at the tip of the red giant branch (with typical radius $10^{-2}R_\odot$) that we make use of when calculating constraints in Section~\ref{sec:RG} are shown in Fig.~\ref{Fig: RG profiles}.

\section{Matrix element for the pion process}\label{app:pion}

Here we give expressions for the matrix elements for KK graviton production from pions, corresponding to the Feynman diagrams in Figure~\ref{fig:pion diagrams}. With the conventions for the momenta in that figure, we have
\begin{align}
M^s_{\alpha\beta} &= \frac{i \kappa f_{pp}}{4\sqrt{2}m_\pi}
   \frac{1}{(p_1+p_2)^2-m_N^2}
   \left[(p_1+p_2+p_4)^\sigma(\gamma_\alpha \eta_{\beta\sigma}
   +\gamma_\beta\eta_{\alpha\sigma}-2\gamma_\sigma\eta_{\alpha\beta})
   +4m_N\eta_{\alpha\beta}\right]
   \\ 
   &\hspace{140pt} \times (\slashed{p_1}+\slashed{p_2}+m_N)\slashed{p_1}\gamma^5
   \nonumber
\\[2ex]
M^u_{\alpha\beta} &= \frac{i \kappa f_{pp}}{4\sqrt{2}m_\pi}
   \frac{1}{(p_2-p_3)^2-m_N^2}
   \slashed{p_1}\gamma^5(\slashed{p_2}-\slashed{p_3}+m_N)
   \\ 
   & \hspace{140pt} \times \left[(2p_2-p_3)^\sigma(\gamma_\alpha \eta_{\beta\sigma}
   +\gamma_\beta\eta_{\alpha\sigma}-2\gamma_\sigma\eta_{\alpha\beta})
   +4m_N\eta_{\alpha\beta}\right]
   \nonumber
\\[2ex]
M^t_{\alpha\beta} &= \frac{i\kappa f_{pp}}{\sqrt{2}m_\pi}
   \frac{1}{(p_1-p_3)^2-m_\pi^2}
   (\slashed{p_1}-\slashed{p_3})\gamma^5
   \left[p_1^\rho(p_1-p_3)^\sigma C_{\alpha \beta \rho \sigma}
   + m_\pi^2 \eta_{\alpha \beta} \right]
\\[2ex]
M^c_{\alpha\beta} &= \frac{-i\kappa f_{pp}}{2\sqrt{2}m_\pi}
   p_1^\sigma(\gamma_\alpha \eta_{\beta\sigma}
   +\gamma_\beta\eta_{\alpha\sigma}-2\gamma_\sigma\eta_{\alpha\beta})\gamma^5, \label{eq:contact}
\end{align}
where $C_{\mu\nu\rho\sigma}$ is defined in Ref.~\cite{Han:1998sg}. Here $f_{pp} = g_A m_\pi/(2F_\pi) \approx1$ with $g_A\approx1.3$ the axial-vector coupling constant of pions and nucleons and $F_\pi \approx 93\MeV$ is the pion decay constant.

\section{Decay channels}\label{app:decay}

In this Appendix we count the number of channels by which a KK graviton of mass $m_h\gg m_{\rm KK}$ can decay to two lighter KK gravitons, assuming that the total KK number can change by up to $\delta n$ during the decay, as described in Section~\ref{sec:decay}. 

Consider first the case of one extra dimension. We label the KK number of an initial state $N \equiv m_h/m_{\rm KK}$ (we take this to be positive without loss of generality), and those of the final state, $a$ and $b$. The number of decay channels is given by the number of integers $a$, $b$ such that
\beq
N-\delta n < a+b < N~,
\eeq
where the lower bound is from the maximum violation of KK number, and the upper bound is from conservation of energy. Hence, the number of channels is of order 
\beq
N \delta n = \frac{m_h}{m_{\rm KK}} \delta n~,
\eeq
(we are not concerned with order-one numerical factors).

Next we consider the case of two equal radius extra dimensions, assuming violation of up to $\delta n$ is allowed for the KK numbers associated with both compactified directions. Without loss of generality, we take the KK numbers of the initial state to be $(N,0)$, where $N=m_h/m_{\rm KK}$ as before. Call the KK numbers of the final states ${\bf a}=(a_1,a_2)$ and ${\bf b}=(b_1,b_2)$. We must count the number of pairs of vectors over the integers $({\bf a},{\bf b})$ that satisfy 
\beq \label{eq:aba}
N-\delta n< a_1 + b_1< N+\delta n \, , \quad {\rm and}\quad -\delta n< a_2 + b_2 < \delta n ~,
\eeq
(from KK number) and
\beq
|{\bf a}|+|{\bf b}| < N~,
\eeq
from energy conservation. These conditions are equivalent to
\beq \label{eq:ab}
|{\bf a}|+|{\bf f}-{\bf a}| < N~,
\eeq
where ${\bf f}\equiv(f_1,f_2)={\bf a}+{\bf b}$ is such that $f_1\in (N-\delta n,N+\delta n)$ and $f_2\in (-\delta n,\delta n)$. 

Consider a particular fixed ${\bf f}=(N-\delta n_1,\delta n_2)$. Eq.~\eqref{eq:ab} corresponds to points inside an ellipse with foci at ${\bf 0}$ and ${\bf f}$ and major axis of length $N$. The number of such points is the area of the ellipse, which (dropping numerical factors) is
\beq
N \sqrt{N^2-|{\bf f}|^2} \simeq N \sqrt{N \delta n_1} ~, 
\eeq
where we used $\delta n_1 \leq \delta n\ll N$. 

Since the number of different ${\bf f}$ is of order $\delta n^2$, the total number of open decay channels is
\beq \label{eq:c2}
\delta n^2 N \sqrt{N \delta n} = \left( \frac{m_h}{m_{\rm KK}} \right)^{3/2} \delta n^{5/2} ~,
\eeq
where we neglect an overall numerical factor that is about $0.3$. Eq.~\eqref{eq:c2} is the expression used in the main text.

\bibliographystyle{JHEP}
\bibliography{biblio}

\end{document}